\documentclass[11pt,a4paper]{article}
\usepackage{url}
\usepackage{jcappub}
\usepackage{amsfonts}
\usepackage{subfigure}
\usepackage{graphicx}
\usepackage{appendix}

\usepackage{epsfig}  
\usepackage{graphicx}   
\usepackage{slashed}       
\usepackage{tikz}
\usepackage{url}
\usepackage{color}
\usepackage{multirow}
\usepackage{comment}
\usepackage{amssymb}
\usepackage[capitalize]{cleveref}
\usepackage{bm}
\usepackage{soul}

\DeclareMathOperator{\cm}{cm}
\DeclareMathOperator{\GeV}{GeV}

\DeclareMathOperator{\MeV}{MeV}

\DeclareMathOperator{\kpc}{kpc}

\newcommand{\beq}{\begin{equation}}
\newcommand{\eeq}{\end{equation}}

\newcommand{\gag}{g_{a\gamma}}
\newcommand{\dSN}{R_{\text{SN}}}
\newcommand{\diff}{\mathrm{d}}
\newcommand{\alpSpectrum}{\frac{\diff N_a}{\diff \omega_a}}
\newcommand{\photonSpectralFlux}{\frac{\diff \Phi_\gamma}{\diff \omega_\gamma}}
\newcommand{\lat}[0]{\textit{Fermi}-LAT}

\crefname{pluralequation}{Eqs.}{Eqs.}
\creflabelformat{pluralequation}{(#2#1#3)}

\allowdisplaybreaks

\bibpunct{[}{]}{,}{n}{}{,}

\definecolor{forestgreen}{rgb}{0.13, 0.545, 0.13}

\makeatletter
\gdef\@fpheader{}
\makeatother

\begin{document}

\title{Investigating the gamma-ray burst from decaying MeV-scale axion-like particles produced in supernova explosions}

\author[a]{Eike M\"uller,}
\emailAdd{eike.muller@fysik.su.se}

\author[b]{Francesca Calore,}
\emailAdd{calore@lapth.cnrs.fr}

\author[a]{Pierluca Carenza,}
\emailAdd{pierluca.carenza@fysik.su.se}

\author[b,c]{Christopher Eckner,}
 \emailAdd{eckner@lapth.cnrs.fr}

\author[a]{M.C.~David Marsh}
\emailAdd{david.marsh@fysik.su.se}

\affiliation[a]{The Oskar Klein Centre, Department of Physics, Stockholm University, 10691 Stockholm, Sweden}
\affiliation[b]{LAPTh, CNRS,  USMB, F-74940 Annecy, France}
\affiliation[c]{LAPP, CNRS,  USMB, F-74940 Annecy, France}

\abstract{
    We investigate the characteristics of the gamma-ray signal following the decay of MeV-scale Axion-Like Particles (ALPs) coupled to photons which are produced in a Supernova (SN) explosion.
    This analysis is the first to include the production of heavier ALPs through the photon coalescence process, enlarging the mass range of ALPs that could be observed in this way and giving a stronger bound from the observation of SN 1987A.
    Furthermore, we present a new analytical method for calculating the predicted gamma-ray signal from ALP decays. With this method we can rigorously prove the validity of an approximation that has been used in some of the previous literature, which we show here to be valid only if all gamma rays arrive under extremely small observation angles (i.e.~very close to the line of sight to the SN). However, it also shows where the approximation is not valid, and offers an efficient alternative to calculate the ALP-induced gamma-ray flux in a general setting when the observation angles are not guaranteed to be small.
    We also estimate the sensitivity of the \textit{Fermi} Large Area Telescope (\lat) to this gamma-ray signal from a future nearby SN and show that in the case of a non-observation the current bounds on the ALP-photon coupling $ \gag $ are strengthened by about an order of magnitude.
    In the case of an observation, we show that it may be possible to reconstruct the product $ \gag^2 m_a $, with $ m_a $ the mass of the ALP.
}
\maketitle

\section{Introduction}

The study of heavy Axion-Like Particles (ALPs) in the mass range between keV and MeV is of great interest due to their significant impact in cosmology and astrophysics. For instance, MeV-scale ALPs can leave an imprint on, e.g., Big Bang Nucleosynthesis and the Cosmic Microwave Background~\cite{Cadamuro:2010cz,Cadamuro:2011fd,Depta:2020wmr,Balazs:2022tjl}, the evolution of low-mass stars~\cite{Raffelt:1987yu,Carenza:2020zil,Dolan:2021rya,Lucente:2022wai}, or the explosion energy of low-energy Supernovae (SNe)~\cite{Sung:2019xie,Caputo:2022mah}. Moreover, this mass range is accessible to colliders and beam-dump experiments, providing a promising avenue for massive ALP detection~\cite{Jaeckel:2015jla,Dolan:2017osp,Dobrich:2019dxc,Banerjee:2020fue}.

Recently, there has been renewed interest in studying the gamma-ray sky in the MeV range~\cite{2021ExA....51.1225D}, and gamma-ray observations in this energy range can be used to probe MeV-scale ALPs produced in SNe. Indeed, besides providing an additional cooling channel for the SN core that could shorten the neutrino burst \cite{Lucente:2020whw}, heavy ALPs have a high decay rate into photons leading to a burst of gamma rays as ALPs escape a SN and undergo decay. 
Based on this emission process, a constraint on the ALP-photon coupling was established in Refs.~\cite{Giannotti:2010ty,Jaeckel:2017tud,Hoof:2022xbe} exploiting the non-observation of a gamma-ray burst by the Solar Maximum Mission (SMM) up to 223 seconds after the arrival and measurement of the first neutrinos from SN 1987A, see also Refs.~\cite{Oberauer:1993yr,Jaffe:1995sw,Brockway:1996yr,Grifols:1996id,Payez:2014xsa}. The calculation in Ref.~\cite{Jaeckel:2017tud} was performed using a Monte Carlo approach to generate the ALP decay events and their corresponding gamma-ray signals. Recently, this method was improved and simplified through the use of analytical techniques, and extended to encompass non-instantaneous ALP emission~\cite{Hoof:2022xbe}. This refinement and extension of the previous calculation is a first step towards a more comprehensive understanding of the heavy ALP phenomenology that we intend to continue with this work.

Given the widespread interest in this field, we conduct a thorough revision of the gamma-ray bound on decaying ALPs in this paper, incorporating the appropriate spectrum for massive ALPs produced in a SN through Primakoff conversion and the previously neglected photon coalescence, with the latter being the primary production mechanism for ALPs with masses greater than $70$~MeV.
We show, in some detail, how the photon signal can be calculated given a flux of ALPs originating from a SN explosion. We derive a fully general expression for the differential photon fluence, and use it to rigorously prove the validity of an often used approximation going back to Refs.~\cite{Oberauer:1993yr,Jaffe:1995sw}. With this approximation we can very efficiently calculate the expected gamma-ray signal from decaying ALPs, and show that the resulting bound from SN 1987A extends up to $ m_a \sim 280 $~MeV. Furthermore, we explore the capability of \lat~to detect a gamma-ray burst induced by ALPs from a future nearby SN, similar to what has been done previously for light ALPs that convert into photons in the galactic magnetic field \cite{Crnogorcevic:2021wyj}. Our updated analysis provides a comprehensive and precise understanding of the topic and its associated phenomenology, offering insights into the detection of heavy ALPs.

A recent comprehensive study explored the possibility of reconstructing the ALP properties based on a future detection of solar ALPs with sub-eV masses in IAXO~\cite{Hoof:2021mld}. Their findings indicate that ALP models can typically be well differentiated, and the well-understood characteristics of the Sun can even be leveraged to use ALPs as probes for astrophysical investigations. Similarly, in the event of a signal from a SN, it may be possible to explore the MeV mass range and determine the extent to which ALP properties can be reconstructed. We extensively discuss this possibility for a potential observation of such a signal by \lat~and find that the photon coupling and mass of the ALP cannot be inferred independently from each other. 

In \cref{sec:production} we discuss the ALP production in a SN, highlighting the importance of photon coalescence in the case of heavy ALPs. In \cref{sec:fluence}, we analyze the gamma-ray signal induced by ALP decay and in  \cref{sec:SN 1987A_bound} we use this calculation to revisit constraints from SN 1987A based on the SMM observations.
The possibility of reconstructing ALP properties after an observation of a future nearby SN event is elaborated in \cref{sec:futureSN}. Finally, in \cref{sec:conclusions}, we summarize and conclude.

Throughout this work we set $ c = \hbar = k_{\text{B}} = 1 $.

\section{Production of heavy ALPs in supernovae}
\label{sec:production}
In the hot and dense plasma at the core of a SN, even weakly interacting particles like ALPs will be produced in large amounts. Here, we only assume a coupling between ALPs and photons, in which case the two relevant production process are Primakoff conversion and photon coalescence \cite{Raffelt:1985nk,DiLella:2000dn}.
The Primakoff process, i.e.~the conversion of a photon into an ALP in the electrostatic field of charged particles in the plasma, has the following production spectrum, i.e.~a spectral rate of change of the number density, in the case of massive ALPs~\cite{DiLella:2000dn,Carenza:2020zil}:
\begin{equation} \label{eq:PrimakoffSpectralProductionRate}
\begin{aligned}
    \dfrac{\diff^2 n_a^{\text{Prim.}}}{\diff t_{\text{pb}}\, \diff\omega_{a}} = \gag^2 \dfrac{T\kappa_s^2}{32\pi^{3}} \frac{k \, p_a}{e^{\omega_a / T} - 1} \Biggl\{ &\dfrac{\left[\left(k+p_a\right)^2+\kappa_s^2\right]\left[\left(k-p_a\right)^2+\kappa_s^2\right]}{4kp_a\kappa_s^2}\ln\left[\dfrac{(k+p_a)^2+\kappa_s^2}{(k-p_a)^2+\kappa_s^2}\right]\\
    &\quad - \dfrac{\left(k^2-p_a^2\right)^2}{4kp_a\kappa_s^2}\ln\left[\dfrac{(k+p_a)^2}{(k-p_a)^2}\right]-1 \Biggr\}\,,
\end{aligned}
\end{equation}
where $ \omega_a $ is the energy of the emitted ALP and $ t_{\text{pb}} $ is the time after the SN core bounce, both measured in the local frame of the ALP, $p_a=\sqrt{\omega_a^2-m_a^2}$ and $k=\sqrt{\omega_a^2-{\omega_{\rm pl}}^2}$ are the ALP and photon momentum respectively, $T$ is the temperature of the plasma, $\omega_{\rm pl}$ is the plasma frequency, and $\kappa_s$ the screening scale for a degenerate nucleon gas~\cite{Lucente:2020whw,Payez:2014xsa}. The photon and ALP energy are identical since we can neglect the recoil of the heavy proton. The Primakoff process mainly takes place in the electric field of protons since electrons are strongly degenerate \cite{Payez:2014xsa}.

It is well known \cite{DiLella:2000dn,Lucente:2020whw} that in a SN core ALPs with masses $ m_a \geq 2 \omega_{\text{pl}} \sim \mathcal{O}(25 \text{ MeV}) $ can also be efficiently produced by the inverse decay $ \gamma \gamma \to a $, often called photon coalescence. However, in the analysis of the gamma-ray bound on the photon coupling of ALPs \cite{Giannotti:2010ty,Jaeckel:2017tud,Caputo:2021rux,Balazs:2022tjl,Hoof:2022xbe}, this production process has been omitted so far. In fact, most of the cited works use the same ALP spectrum that was found in Ref.~\cite{Payez:2014xsa} where photon coalescence was not included since only ultralight ALPs were studied. Here, we improve on the previous literature by also taking this process into account.
The spectral production rate for photon coalescence including a non-zero effective photon mass and quantum statistics was derived in Ref.~\cite{Ferreira:2022xlw}:
\begin{equation} \label{eq:photonCoalescenceSpectralProductionRate}
    \frac{\diff^2 n_a^{\text{PC}}}{\diff t_{\text{pb}} \, \diff\omega_a} = \frac{\gag^2 m_a^4}{128 \pi^3} \left(1 - \frac{4 \omega_{\text{pl}}^2}{m_a^2}\right) 
    \int_{\omega_{\text{min}}}^{\omega_{\text{max}}} \diff \omega_\gamma \, \left[ \left(e^{\omega_\gamma/T} - 1\right) \left(e^{(\omega_a-\omega_\gamma)/T} - 1\right) \right]^{-1} \, ,
\end{equation}
with the minimal and maximal photon energy
\begin{equation}
    \omega_{\text{min,max}} = \frac{1}{2} \left( \omega_a \mp p_a \sqrt{1 - \frac{4 \omega_{\text{pl}}^2}{m_a^2}} \right) \, .
\end{equation}

The ALP spectrum $ \alpSpectrum $ is the volume and time-integral of the total spectral production rate (i.e.~the sum of Primakoff and photon coalescence contributions in \cref{eq:PrimakoffSpectralProductionRate,eq:photonCoalescenceSpectralProductionRate}). The quantities in \cref{eq:PrimakoffSpectralProductionRate,eq:photonCoalescenceSpectralProductionRate} that depend on radial distance to the SN center, $ r $, and time after the core bounce, $ t_{\text{pb}} $, are the temperature $ T(r,t_{\text{pb}}) $, the screening scale $ \kappa_S(r,t_{\text{pb}}) $ and the plasma frequency $ \omega_{\text{pl}}(r,t_{\text{pb}}) $. Following the procedure in Ref.~\cite{Ferreira:2022xlw}, we take their tabulated values at different radii and times from the one-dimensional, numerical SN model described in Ref.~\cite{Fischer:2021jfm}.\footnote{We use the \textit{reference model} of Ref.~\cite{Fischer:2021jfm} that does not include any additional cooling or energy transfer by ALPs since we are considering such small couplings here that their effect on the explosion dynamics is negligible.} For practicality, we cut the radial integral off at $ R_\text{max} = 24 $~km since the contribution from high radii and therefore small temperatures and densities is negligible, and we cut the time integral off at $ t_{\text{pb}}^{\text{min}} = 0.5 $~s since our SN profiles are not smooth before that time. Furthermore, it was recently shown in Ref.~\cite{Hoof:2022xbe} that the production of ALPs is well approximated as instantaneous, i.e.~that the dependence of the resulting gamma-ray signal on the delay-time $t$ can be neglected (if this were not the case, we should not integrate over $ t_{\text{pb}} $ and use the time-dependent ALP spectrum \cite{Hoof:2022xbe}).
With that we find the following ALP spectrum:
\begin{equation}
\begin{split} \label{eq:alpSpectrum}
    \alpSpectrum(\omega_a) &= 4\pi \int \diff t_{\text{pb}} \, \diff r \, r^2 \, \ell^{-1}(r,t_{\text{pb}}) \dfrac{\diff^2 n_a}{\diff t_{\text{pb}}\,\diff\omega_{a}^{\text{loc}}}(r,t_{\text{pb}},\ell^{-1}(r,t_{\text{pb}}) \, \omega_a) \, ,
\end{split}
\end{equation}
where, from here on, $ \omega_a $ is the energy of the ALP in the frame of an observer far away from the SN,\footnote{Such an observer far away from the SN will in fact never measure a spectrum as shown in \cref{eq:alpSpectrum} in a finite time interval since all ALPs have slightly different velocities and hence the burst of ALPs will disperse. This effect is taken into account when we calculate the observable photon spectrum below.} which is red-shifted compared to the local energy with which it is produced in the SN core $ \omega_a^{\text{loc}} = \ell^{-1}(r,t_{\text{pb}}) \, \omega_a $; here, $ \ell(r,t_{\text{pb}}) $ is called the lapse function, which we also take from the numerical SN model in Ref.~\cite{Fischer:2021jfm}. Note that the spectral production rates in \cref{eq:PrimakoffSpectralProductionRate,eq:photonCoalescenceSpectralProductionRate} are written as derivatives and functions of \emph{local} energies, but we have suppressed the superscript ``loc'' for notational simplicity.

\begin{figure}
    \centering
    \includegraphics[width=.99\textwidth]{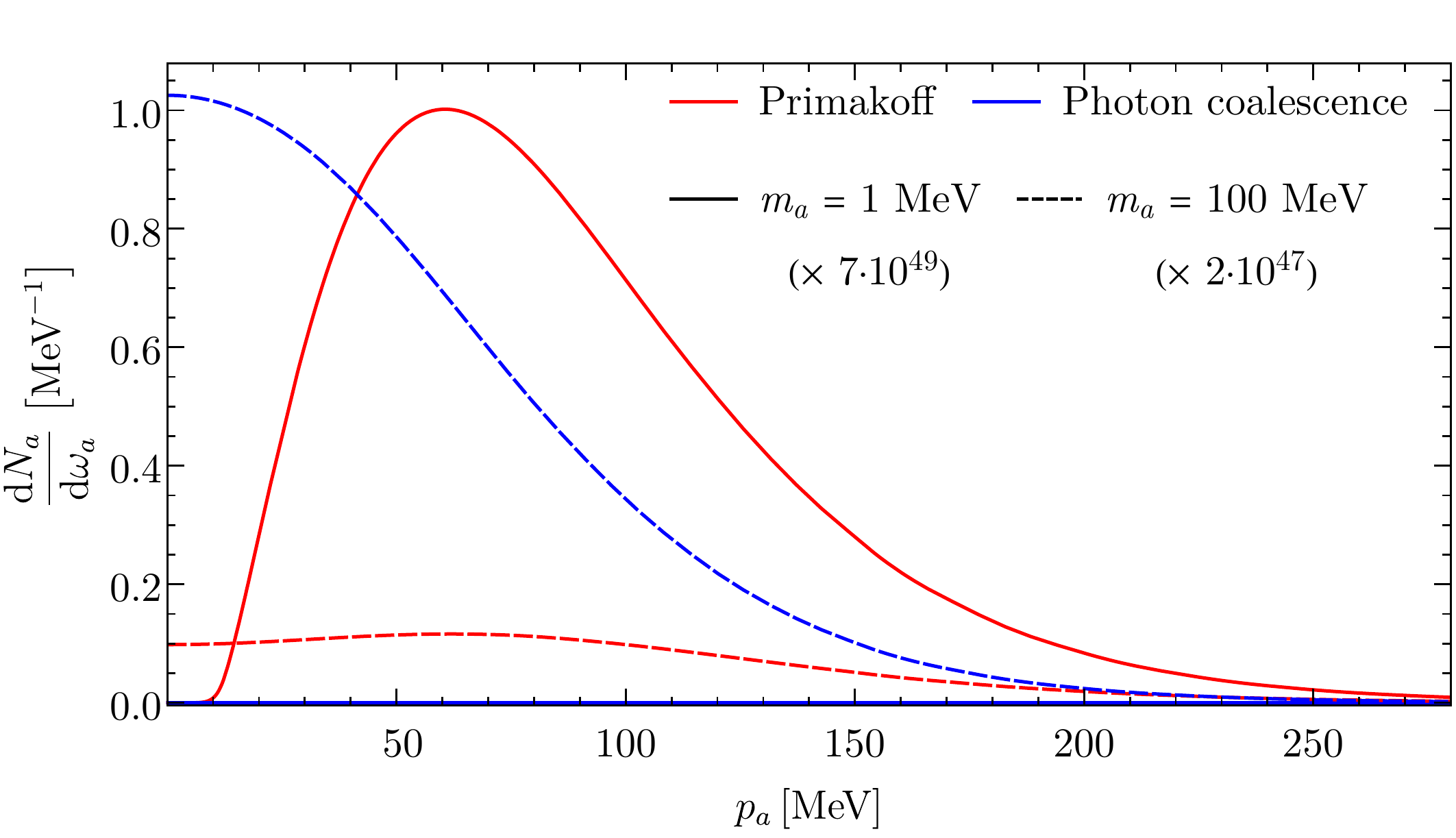}
    \caption{Primakoff (red) and photon coalescence (blue) spectra of ALPs with $  m_a = 1 $~MeV and $ \gag = 10^{-11} \text{ GeV}^{-1} $ (solid), and $  m_a = 100 $~MeV and $ \gag = 4 \cdot 10^{-13} \text{ GeV}^{-1} $ (dashed) produced in our SN model. The two choices of parameters correspond to \textbf{case 1} and \textbf{case 2} discussed in \cref{sec:FermiFit}. Note that we show the spectrum as a function of momentum for ease of presentation, even though it is a differential with respect to the energy.
    }
    \label{fig:ALPspectra}
\end{figure}
In \cref{fig:ALPspectra} we show the Primakoff and photon coalescence spectra in red and blue, respectively, for light, $ m_a = 1 $~MeV (solid lines), and heavy, $ m_a = 100 $~MeV (dashed lines), ALPs. The effect of the gravitational red-shift can clearly be seen for the heavier ALPs: the spectrum does not go to zero for $ p_a \to 0 $ because ALPs that are produced with a low local momentum are trapped in the SN and do not escape, while the momenta of the slowest ALPs that do escape are then red-shifted towards $ p_a \to 0 $ for a distant observer.
It also becomes clear from \cref{fig:ALPspectra} that for heavy ALPs photon coalescence is the dominant production process in a SN plasma, with a production spectrum about a factor of 100 larger than that of the Primakoff effect for $ m_a = 100 $~MeV. The total number of ALPs produced by photon coalescence is equal to that of ALPs produced by the Primakoff effect for a mass of $ m_a = 70 $~MeV, and is larger for all heavier masses. As we will show in \cref{sec:SN 1987A_bound}, this not only slightly strengthens the bound whenever photon coalescence is kinematically possible but especially also extends the bound to higher masses. It is therefore important for an accurate bound on the ALP parameter space to include the photon coalescence contribution to the ALP production spectrum.

\section{The gamma-ray fluence from decaying SN ALPs} \label{sec:fluence}
The temperatures and densities reached in the plasma of a SN explosion are high enough to produce astrophysically relevant amounts of weakly coupled ALPs with masses up to hundreds of MeV. If these ALPs are coupled weakly enough, they can escape the plasma of the SN core. Heavy ALPs (here we consider masses of 10~keV and above) will then eventually decay into a pair of photons, if this is their only coupling as we will assume throughout this work. If enough of those photons reach Earth, they would be observable as a gamma-ray signal \cite{Giannotti:2010ty,Jaeckel:2017tud}.

In the following, we will describe how the differential fluence of these gamma-ray photons, given a spectrum of ALPs produced in the SN, can be calculated.

\subsection{General formula for the differential fluence}
A heavy ALP emitted from a SN travels a distance $ L $ before decaying into a pair of photons of which one might reach a detector near Earth. Since the initial ALP flux is assumed to be isotropic and radial, the resulting flux of photons is spherically symmetric. At any distance $ \dSN $ from the SN, where we assume an observer is located, the total (i.e.~time-integrated) surface density of gamma rays, the so-called fluence, is
\begin{equation}
    F_\gamma^{\text{tot}} = \frac{N_\gamma^{\text{tot}}}{4\pi \dSN^2} \, ,
\end{equation}
where the denominator is the area of the sphere with radius $ \dSN $, and $ N_\gamma^{\text{tot}} \leq 2 N_a $ is the total number of gamma rays reaching this radius; it is smaller than twice (two photons originate from one ALP decay) the total number of produced ALPs, $ N_a $, because not all ALPs have necessarily decayed at the radius $ \dSN $. From the ALPs that decay at larger radii only those photons that are emitted (sufficiently) backwards during the decay can reach the sphere. We can calculate $ N_\gamma^{\text{tot}} $ (or rather its expectation value) by integrating the differential ALP spectrum $ \alpSpectrum $ multiplied with the probability $ P_\gamma^{\text{obs}} $ that a photon produced in the decay reaches the radius $ \dSN $, and is thus in principle observable:
\begin{equation}
    N_\gamma = 2 \int_{m_a}^{\infty} \diff \omega_a \, P^{\text{obs}}_\gamma (\omega_a) \, \alpSpectrum \, ,
\end{equation}
where $ \omega_a $ is the energy of the ALP.
\begin{figure}
    \centering
    \includegraphics[width=.95\textwidth]{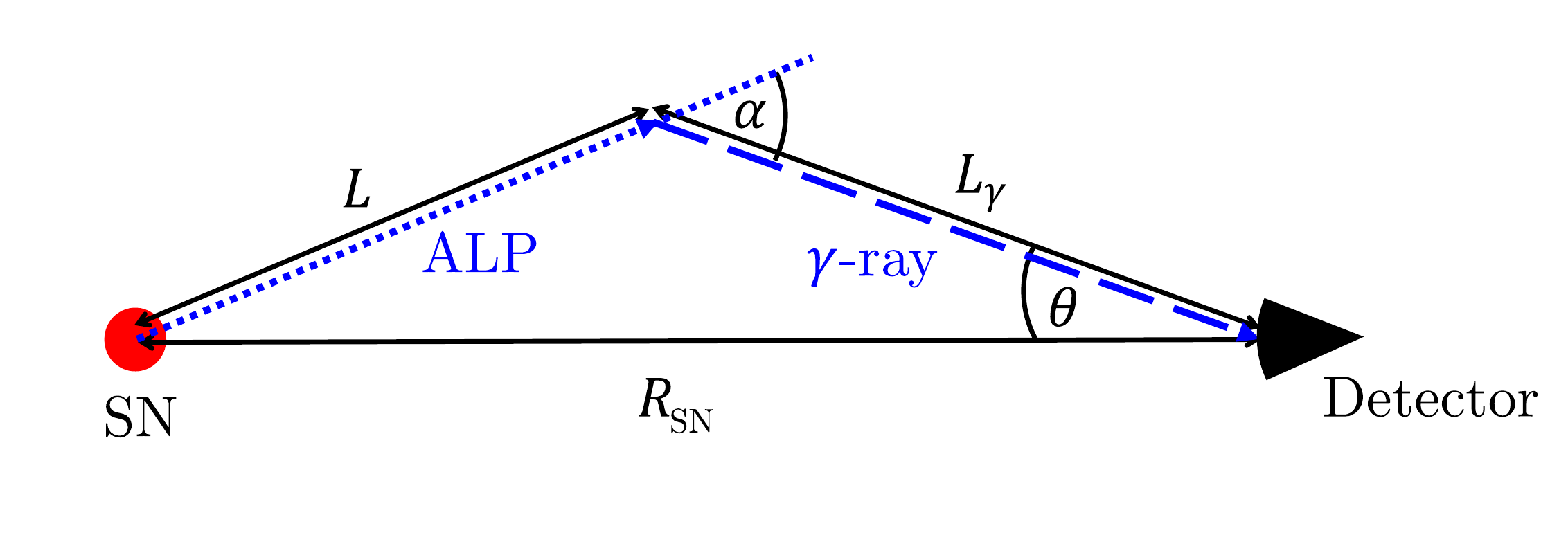}
    \caption{Geometry of the ALP to photon decay. Note that only the photon reaching Earth is shown.}
    \label{fig:decayGeometry}
\end{figure}
The probability $ P_\gamma^{\text{obs}} $ is the integral of the product of the distribution functions for ALP travel length $ L $ and the angle between the ALP and photon momenta $ \alpha $, as shown in \cref{fig:decayGeometry}:
\begin{gather}
    P^{\text{obs}}_\gamma (\omega_a) = \int \diff c_\alpha \int \diff L \, f_{c_\alpha}(c_\alpha, \omega_a) \, f_L(L, \omega_a) \, \Theta_{\text{cons.}}(\omega_a, c_\alpha, L) \, , \\
    \text{with }
    f_L(L, \omega_a) = \ell_a^{-1}(\omega_a) \exp(- L / \ell_a(\omega_a)) \, ,
    \label{eq:fL} \\
    f_{c_\alpha}(c_\alpha, \omega_a) = \frac{m_a^2}{2 \omega_a^2 (1 - c_\alpha \beta_a)^2} \, , \label{eq:fc}
\end{gather}
where $ c_\alpha \equiv \cos(\alpha) $, $ \beta_a = p_a / \omega_a $ the relativistic velocity of the ALP, and $ p_a = \sqrt{\omega_a^2 - m_a^2} $ its momentum. The decay length $ \ell_a $ of an ALP only coupled to photons with energy $ \omega_a $ in the rest frame of Earth and SN is
\begin{equation}
    \ell_a(\omega_a) = \frac{\gamma_a \beta_a}{\Gamma_a^0} \equiv \frac{\omega_a}{m_a} \sqrt{1 - \frac{m_a^2}{\omega_a^2}} \frac{64 \pi}{\gag^2 m_a^3},
\end{equation}
where $ \gamma_a = \frac{\omega_a}{m_a} $, and $ \Gamma_a^0 $ is the decay rate of the ALP in its rest frame. The decay-angle distribution $ f_{c_\alpha} $ is obtained from an isotropic distribution in the ALP's rest frame, boosted into the Earth-SN rest frame~\cite{Landau:1975pou}.
The probability $ P^{\text{obs}}_\gamma $ is smaller than 1 because of the constraints that we have to put on the geometric and kinematic quantities in the form of a product of Heaviside functions that we call $ \Theta_{\text{cons.}}(\omega_a, c_\alpha, L) $.
To determine $ F_\gamma^{\text{tot}} $ we consider all the constraints that prevent photons from reaching the radius $ \dSN $. 
The triangle shown in \cref{fig:decayGeometry} can be constructed if and only if
\begin{equation} \label{eq:geometricConditions}
    \Theta_{\text{cons.}}(\omega_a, c_\alpha, L) \supset \Theta \left( \frac{\dSN}{\sqrt{1 - c_\alpha^2}} - L \right) \cdot \Theta \left( c_\theta^\pm - c_\alpha \frac{L}{\dSN} \right) \, ,
\end{equation}
where
\begin{equation} \label{eq:cosThetaFunction}
  c_\theta^\pm =  c_\theta^\pm(L,c_\alpha) = \pm \sqrt{1 - \left( 1 - c_\alpha^2 \right) \frac{L^2}{\dSN^2}}
\end{equation}
is the cosine of the angle under which the incoming photon will be observed relative to the line of sight to the SN, see \cref{fig:decayGeometry}.
Whenever both solutions for $ c_\theta $ are allowed by the constraint, that part of the integration region has to be counted twice, i.e.~$ \Theta( c_\theta^\pm - x) \equiv \Theta( c_\theta^{+} - x) + \Theta( c_\theta^{-} - x) $. However, most of the contribution with observation angles $ \theta > \pi/2 $ (i.e.~negative $ c_\theta $) is negligible as we will show in \cref{subsec:smallAngle}. Indeed, this configuration is only realized when $ L > \dSN $ and $ c_\alpha < 0 $, i.e.~for ALP decays emitting backward photons.\footnote{The constraints in \cref{eq:geometricConditions} are not mentioned in Ref.~\cite{Hoof:2022xbe}, but are the reason for which the $ t_a^{+} $-solution in Eq.~(B.9) of that reference (which is dropped due to its asymptotic behavior) does indeed often not correspond to a possible geometrical configuration. It could in principle contribute to the photon flux, but $ t_a^{+} $ is typically much larger than the lifetime of heavier ALPs for couplings that are phenomenologically relevant; on the other hand, lighter ALPs, for which decay lengths comparable to $ \dSN $ are possible, do not decay backwards at the relevant energies. Thus, this solution does in practice not contribute to the SN 1987A bound that is examined in Ref.~\cite{Hoof:2022xbe}, as was also discussed in Ref.~\cite{Jaeckel:2017tud}. It could, however, become relevant when time delays much larger than $ \mathcal{O}(100 \text{ s}) $ are observed.}
Furthermore, photons will not reach the Earth if the parent ALP decays inside the SN photosphere, where photons are absorbed by the plasma. This yields a lower bound on $ L $:
\begin{equation} \label{eq:escapeCondition}
    \Theta_{\text{cons.}}(\omega_a, c_\alpha, L) \supset \Theta(L - R_*) \, ,
\end{equation}
where $ R_* \sim \mathcal{O}(10^{12})$~cm is the radius of the SN photosphere~\cite{Kazanas:2014mca}. Absorption by the SN plasma can also happen if the ALP initially traveled away from Earth and then decayed backwards, with a resulting photon trajectory that leads through the SN. Since those photons will not be observed, one should also impose
\begin{equation}
    \sin(\theta) > \frac{R_*}{\dSN} \quad \text{ if } c_\alpha < -\sin(\theta) \, .
\end{equation}
This constraint could play a role for heavy ALPs with short decay lengths that decay nearly isotropically right after leaving the SN so that up to $ 50 \% $ of the photons could be absorbed. However, we have checked that this is numerically irrelevant: These ALPs are so short-lived that their fluence is suppressed exponentially as $ \exp(- R_* / \ell_a) $ (according to \cref{eq:escapeCondition}), so that a small increase of the ALP decay length can compensate for the $ \mathcal{O}(10\%) $ absorbed photons, leading to a numerically negligible shift in the $ (\gag, m_a) $ space. Therefore we ignore this constraint in the rest of this work.

Finally, we are typically not interested in the \textit{total} fluence of gamma rays $ F_\gamma^{\text{tot}} $, but instead we want to know the fluence of actually observable gamma rays $ F_\gamma $, taking into account that any given detector is only sensitive to a specific energy range and will only measure for a finite time. To calculate this quantity, $ \Theta_{\text{cons.}}(\omega_a, c_\alpha, L) $ has to be expanded to include also limits on the photon energy and the time at which the photon arrives at the detector (and possibly other observational constraints, e.g.~the angular size of the photon signal). We will discuss these limits in \cref{sec:SN 1987A_bound} when we discuss specific instruments.

Putting everything together, the observable gamma-ray fluence can be written as
\begin{equation}
\begin{aligned} \label{eq:totalFluenceALPVars}
    F_\gamma
    &\equiv \int_{m_a}^{\infty} \mathrm{d} \omega_a \int_{-1}^{1} \mathrm{d}c_\alpha \int_{0}^{\infty} \mathrm{d}L \, \frac{\diff F_\gamma}{\diff \omega_a \, \diff c_\alpha \, \diff L}\\
    &= \frac{2}{4\pi \, \dSN^2} \int_{m_a}^{\infty} \mathrm{d} \omega_a \int_{-1}^{1} \mathrm{d}c_\alpha \int_{0}^{\infty} \mathrm{d}L \, \alpSpectrum f_{c_\alpha}(c_\alpha, \omega_a) \frac{e^{-L/\ell_a(\omega_a)}}{\ell_a(\omega_a)} \Theta_{\text{cons.}}(\omega_a, c_\alpha, L) \, .
\end{aligned}
\end{equation}
This approach was used in the previous literature, e.g.~in Refs.~\cite{Jaeckel:2017tud,Balazs:2022tjl,Ferreira:2022xlw,Hoof:2022xbe}, by either evaluating the integral partly analytically or numerically, or by Monte Carlo simulations. Furthermore, there are approximations to the expression in \cref{eq:totalFluenceALPVars}, e.g.~in Refs.~\cite{Oberauer:1993yr,Jaffe:1995sw,Caputo:2021rux}. In \cref{subsec:observerVariables}, we present a new method to calculate the fluence, which we think is useful to not only rigorously prove these approximations and their range of applicability (\cref{subsec:smallAngle}), but also to calculate the ALP-induced \textit{flux} (i.e.~the time-differential of the fluence) of gamma rays that would e.g.~be observable by the \lat~detector (\cref{sec:futureSN}).

\subsection{Observer variables}\label{subsec:observerVariables}
Even though \cref{eq:totalFluenceALPVars} fully describes the gamma-ray flux that we want to study, we can improve our understanding and the numerical efficiency of the calculation by linking the ALP-related variables $ \omega_a, \, c_\alpha, \, L $ with variables describing the observed photon. Namely these are the energy of the observed photon $ \omega_\gamma$, the time delay of the photon compared to the first neutrinos $ t$, and the cosine of the observation angle $ c_\theta $ as in \cref{fig:decayGeometry}. In this section we show that there is a one-to-one matching between these two sets of variables.

Relativistic kinematics fixes the angle $ \alpha $ between an ALP with energy $ \omega_{a} $ and the photon it decays into with an energy $ \omega_\gamma $ to be
\begin{equation}
    c_{\alpha}=\beta_{a}^{-1}\left( 1 - \frac{m_a^2}{2 \omega_a \, \omega_\gamma} \right) = p_a^{-1} \left( \omega_a - \frac{m_a^2}{2 \, \omega_\gamma} \right) \, ,
\end{equation}
in the frame in which Earth and SN are at rest.
The variable transformation $c_\alpha \mapsto \omega_\gamma $ transforms the decay angle distribution in \cref{eq:fc} into the inverse ALP momentum
\begin{equation}
        f_{c_\alpha} \diff c_\alpha = f_{c_\alpha} \left\lvert\frac{\partial c_\alpha}{\partial \omega_\gamma}\right\rvert \diff \omega_\gamma = p_a^{-1} \diff \omega_\gamma \, .
\end{equation}
From the range $ c_\alpha \in [-1,1] $ we can infer $ \omega_\gamma \in [(\omega_a - p_a)/2, (\omega_a + p_a)/2] $.
The remaining transformation $ (\omega_a, \, L) \mapsto (t, \, c_\theta) $ for fixed $ \omega_\gamma $ requires three identities that can be inferred from \cref{fig:decayGeometry} using elementary geometry:
\begin{equation} \label[pluralequation]{eq:lawOfSines}
    \begin{gathered}
        t = \frac{L}{\beta_a} + L_\gamma - \dSN \, ,\\
        L_\gamma = \frac{s_{\alpha - \theta}}{s_\alpha} \, \dSN = \frac{s_\alpha c_\theta - c_\alpha s_\theta}{s_\alpha} \, \dSN = c_\theta \, \dSN - c_\alpha \, L \, ,\\
        L =  \frac{s_\theta}{s_\alpha} \dSN = \sqrt{\frac{1-c_\theta^2}{1-c_\alpha^2}} \dSN \, ,
    \end{gathered}
\end{equation}
where $ t $ is the time delay between arrival of the photon and the detection of the first neutrino,\footnote{We assume that all ALPs are emitted instantaneously from the SN, since the emission is only efficient for a few seconds while the gamma-ray signal typically stretches over much longer times. That this is indeed a good approximation for the SN 1987A bound on ALPs was recently checked and confirmed in Ref.~\cite{Hoof:2022xbe}.} $ L_{\gamma} $ and $ L $ are the distances traveled by the photon and the ALP, respectively, and $ s_x \equiv \sin(x) $. Note that the first geometrical constraint in \cref{eq:geometricConditions} will always be satisfied when we use these equations to define our variables, since in the last line of \cref{eq:lawOfSines} we know that $ c_\theta^2 \leq 1 $. The second geometrical constraint is non-trivial and ensures that $ L_{\gamma} $ is positive.
Solving these equations for $ \omega_a $ and $ L $ yields the variable transformations
\begin{equation} \label[pluralequation]{eq:trafos}
\begin{split}
    \omega_a(\omega_\gamma, t, c_\theta)
    &= \omega_\gamma + \frac{m_a^2}{4\omega_\gamma} \left( 1 + \frac{1 - c_\theta^2}{(t/\dSN + 1 - c_\theta)^2} \right) \, ,\\
    L(\omega_\gamma, t, c_\theta) &= \frac{2\omega_\gamma \, p_a(\omega_\gamma, t, c_\theta)}{m_a^2} \left( \frac{t}{\dSN} + 1 - c_\theta \right) \, \dSN \, ,
\end{split}
\end{equation}
and the following Jacobian:
\begin{equation}
    \left\lvert\frac{\partial(\omega_a, L)}{\partial(c_\theta, t)}\right\rvert = \frac{|c_\theta|}{(t/\dSN + 1 - c_\theta)^2} \, p_a(\omega_\gamma, t, c_\theta) \, .
\end{equation}
In conclusion, the differential fluence becomes
\begin{equation} \label{eq:diffFluenceObsVars}
\begin{aligned}
    \frac{\diff^3 F_\gamma}{\diff \omega_\gamma \diff t \diff c_\theta} &= \left\lvert\frac{\partial c_\alpha}{\partial \omega_\gamma} \cdot \frac{\partial(\omega_a, L)}{\partial(c_\theta, t)}\right\rvert \frac{\mathrm{d}^3 F_\gamma}{\mathrm{d} \omega_a \mathrm{d} c_\alpha \mathrm{d} L}=\\
    &= \frac{2}{\tau_a} \frac{|c_\theta|}{(t/\dSN + 1 - c_\theta)^2} \frac{\alpSpectrum(\omega_a(\omega_\gamma, t, c_\theta))}{4\pi \, \dSN^2} \frac{m_a}{p_a(\omega_\gamma, t, c_\theta)} \\
    &\quad \times \exp\left[ -\frac{\dSN}{\tau_a} \frac{2 \omega_\gamma}{m_a} \left(\frac{t}{\dSN} + 1 - c_\theta\right) \right] \Theta_{\text{cons.}}(\omega_\gamma, t, c_\theta) \, ,
\end{aligned}
\end{equation}
where we have introduced the ALP's lifetime in its rest frame $ \tau_a \equiv \left(\Gamma_a^0\right)^{-1} = \frac{64 \pi}{\gag^2 m_a^3} $.
Remarkably, since the solutions in \cref{eq:trafos} are unique, there is only one possible trajectory of the ALP-photon system if $ \omega_\gamma, t, c_\theta $ are known. This is in contrast to the original ALP-variable approach in \cref{eq:totalFluenceALPVars,eq:geometricConditions}, where $ c_\theta $ is not necessarily single-valued, or the approach of Ref.~\cite{Hoof:2022xbe} where the ALP's travel time $ t_a = \beta_a^{-1} L $ is not uniquely determined.

The second geometrical constraint in \cref{eq:diffFluenceObsVars}, $ c_\theta \, \dSN > c_\alpha \, L $, can be written as a lower bound on $ c_\theta $:
\begin{equation}
    c_\theta > 1 - \frac{t}{\dSN} \left(\frac{m_a}{2\omega_\gamma} \sqrt{\frac{2\dSN}{t} + 1} - 1 \right) \, .
    \label{eq:thetaBound}
\end{equation}
This condition gives an upper bound on the delay time because $ c_{\theta} < 1 $:
\begin{equation}
        t < \frac{2 m_a^2}{4 \omega_\gamma^2 - m_a^2} \, \dSN \quad \text{if } \omega_\gamma > \frac{m_a}{2} \, .
\end{equation}
For photon energies $ \omega_\gamma < m_a/2 $ arbitrarily large time delays are allowed, corresponding to ALPs that travel away from Earth for a correspondingly long time before decaying and producing a backward photon.
Due to the non-linear nature of \cref{eq:trafos}, the constraint $ L(\omega_\gamma,t,c_\theta) > R_* $ in \cref{eq:escapeCondition} does not in general translate to a single bound on either of the integration parameters, and has to be evaluated numerically for any given choice of observer variables.
Finally, the integration variables $ \omega_\gamma, \,  t, \, c_\theta $ can directly be constrained to certain intervals to match limits of the observational set up. Our choice of observer variables makes these types of constraints trivial to implement, which is one of the motivations to use these variables.

\subsection{Small-angle approximation} \label{subsec:smallAngle}
Another important reason for the choice of observer variables will be discussed in this section. If the photon delay time is not too large, using these variables it is easy to prove that the signal will fall within a cone with small observation angle $ \theta \sim 10^{\mathcal{O}(1)} \mathcal{O}\left( \frac{t}{\dSN} \right)$, corresponding, for instance, to $ \theta \lesssim 10^{-11} $ for the SMM observation of SN 1987A. In this limit, \cref{eq:trafos} drastically simplify and we recover an approximation of the gamma-ray signal that has been derived in Refs.~\cite{Oberauer:1993yr,Jaffe:1995sw}, whose validity is discussed at the end of this section.

We begin by noting that \cref{eq:thetaBound} enforces at least $ \theta \lesssim \sqrt{\frac{m_a}{\omega_\gamma}} \left( \frac{t}{\dSN} \right)^{1/4} $ if $ t \ll \dSN $. If we are only interested in a finite range of photon energies, i.e.~$ \omega_\gamma > \omega_\gamma^{\text{min}} > 0 $ and $ t/\dSN $ is small enough, it is the geometrical condition that enforces observation angles much smaller than 1. Intuitively, if we are only interested in time delays much shorter than the time it takes light to travel from the SN to Earth, then the geometric path length of the combined ALP-photon trajectory cannot be much larger than $ \dSN $, and hence the triangle in \cref{fig:decayGeometry} must have a small area. Large observation angles at a given $ t \ll \dSN $ would be allowed by the geometry (a small-area triangle with two points close to Earth), but kinematically the combination of large $ L $ and small $ \beta_a $, necessary for a backwards decay, are only possible for $ \omega_\gamma \to 0 $.
In the following we will hence assume
\begin{equation} \label{eq:smallAngleTimeCondition}
    t \ll \min \left[1, \, \left( \frac{\omega_\gamma}{m_a} \right)^2, \, \frac{m_a^2}{8 \, \omega_\gamma \, \omega_a^{\text{max}}}\right] \dSN \, ,
\end{equation}
so that, due to the first two arguments of the minimum function, the geometric constraint enforces $ \theta \ll 1 $. The last argument of the minimum is necessary for the following proof that $ \theta $ is not only small compared to 1 but even small on the order of $ t / \dSN $, which is necessary to simplify \cref{eq:trafos}. In this last argument, $ \omega_a^{\text{max}} $ is a cut-off such that essentially no ALPs with higher energies are produced in the SN, i.e.~$ \alpSpectrum(\omega_a \geq \omega_a^{\text{max}}) \simeq 0 $. In our model of SN 1987A for instance, we set $ \omega_a^{\text{max}} \sim 1 $~GeV.

From \cref{eq:trafos} we can deduce that for $ \theta \ll 1 $ the maximum of the ALP energy $ \omega_a(\theta) $ is reached for $ \theta^{\text{max}} = \sqrt{2 \frac{t}{\dSN}} $. At this point the ALP energy is
\begin{equation}
    \omega_a(\theta^{\text{max}}) \simeq \frac{m_a^2}{8\omega_\gamma} \frac{\dSN}{t}=1.3\times10^{6}\GeV\left(\frac{m_{a}}{\MeV}\right)^{2}\left(\frac{\omega_{\gamma}}{\MeV}\right)^{-1}\left(\frac{\dSN}{10~\kpc}\right)\left(\frac{{\rm t}}{100~{\rm s}}\right)^{-1} \, ,
\end{equation}
and we need the third argument of the minimum in \cref{eq:smallAngleTimeCondition} to infer that $ \omega_a(\theta^{\text{max}}) \gg \omega_a^{\text{max}} $. Since the number of ALPs is Boltzmann-suppressed at such high energies, there are no observable photons at angles of order $ \theta^{\text{max}} $. Even though $ \omega_a(\theta) $ decreases again for angles larger than $ \theta^{\text{max}} $, it stays above $ \omega_a^{\text{max}} $ as long as the geometric constraint is fulfilled. Therefore, the ALP spectrum enforces
\begin{equation} \label{eq:smallAngleDefinition}
    \theta \lesssim 2 \, \frac{\sqrt{\omega_\gamma \, \omega_a^{\text{max}}}}{m_a} \frac{t}{\dSN} \, ,
\end{equation}
since for such values of $ \theta $ not only is the geometric constraint fulfilled but also $ \omega_a \lesssim \omega_a^{\text{max}} $, again following \cref{eq:trafos}.
This is the case we refer to as the \textit{small angle approximation}. Note that both a geometric and a spectral constraint are necessary to establish the validity of this approximation, and that furthermore the distributions of decay length and angle did not enter the above argument at all.
In the small angle approximation, \cref{eq:trafos} simplify considerably:
\begin{equation} \label[pluralequation]{eq:trafosSmallAngleApproximation}
\begin{gathered}
    \omega_a(\omega_\gamma, t, c_\theta) = \omega_\gamma + \frac{m_a^2}{4\omega_\gamma} \left[ 1 + \left( \frac{\theta}{t/\dSN} \right)^2 \right] \, ,\\
    L(\omega_\gamma, t, c_\theta) = \frac{2\omega_\gamma \, p_a(\omega_\gamma, t, c_\theta)}{m_a^2} \, t \, ,
\end{gathered}
\end{equation}
where we have ignored terms that are small according to \cref{eq:smallAngleTimeCondition,eq:smallAngleDefinition}. Note that $ \theta \ll 1 $ is not a sufficient assumption for this approximation because also $ t/\dSN \ll 1 $.
Importantly, the first line of \cref{eq:trafosSmallAngleApproximation} is uniquely invertible and we can change variables again: $ c_\theta \mapsto \omega_a $. Note that $ \theta $ is so small in the small angle approximation that it is not realistically observable as it will always lie out of reach of the angular resolution of instruments like the gamma-ray spectrometer onboard the SMM satellite or \lat. Hence, transforming back to the ALP energy instead of $ \theta $ as a variable does not contradict the philosophy of observer variables: we have to integrate over whichever variable corresponds to $ \theta $ anyway, and integrating over $ \omega_a $ makes the expression for the fluence simpler since the ALP spectrum depends on $ \omega_a $.

The differential fluence in the small angle approximation can finally be written as
\begin{equation} \label{eq:diffFluenceSmallAngle}
    \frac{\diff^3 F_\gamma}{\diff \omega_\gamma \, \diff t \, \diff \omega_a} = \frac{4}{4 \pi \dSN^2} \frac{\omega_\gamma}{\tau_a \, p_a \, m_a} \alpSpectrum
    \, e^{-\frac{t}{\tau_a} \frac{2\omega_\gamma}{m_a}} \, \Theta_{\text{cons.}}(\omega_\gamma, t, \omega_a) \, .
\end{equation}
This expression agrees with those found in Refs.~\cite{Oberauer:1993yr,Jaffe:1995sw}. However, in those references (and some of the later literature \cite{Raffelt:1996wa,Caputo:2021rux}), \cref{eq:diffFluenceSmallAngle} has been assumed to hold when the decay length of the ALP is short compared to $ \dSN $. However, for the SN 1987A constraint on ALPs with masses below $ m_a \sim 230 $~keV the typical decay length is in fact larger than the distance between Earth and the SN. In this section, we proved that \cref{eq:diffFluenceSmallAngle} is still an excellent description of the photon flux for those masses and couplings, as long as the observation time is small according to \cref{eq:smallAngleTimeCondition}.
Note that the geometric constraint in \cref{eq:thetaBound} is fulfilled by assumption in the small angle approximation. The constraint $ L > R_* $ is easily implemented according to \cref{eq:trafosSmallAngleApproximation}, and can be read as an upper bound on either of the variables $ \omega_\gamma, \, t, \, \omega_a $, when the other two are held constant.
Having derived the number of ALPs produced and the differential fluence of resulting photons arriving on Earth, we can now proceed to study if these photons can be observed.

\section{Gamma-ray bound on ALPs set by SN 1987A}
\label{sec:SN 1987A_bound}

For 223~s after the detection of the SN 1987A neutrino burst, the gamma-ray Spectrometer on board of the Solar Maximum Mission satellite could have observed a gamma-ray burst in direction of the SN. There was no statistically significant excess over the background, and hence we can constrain the interaction between heavy ALPs and photons, as was done in e.g.~\cite{Jaeckel:2017tud,Ferreira:2022xlw,Hoof:2022xbe}. The ALPs produced in the SN are constrained to induce a photon fluence of $ F_\gamma < 1.78 \text{ cm}^{-1} $ at the satellite to be consistent with the (null) observation at the 3 sigma level, or $ F_\gamma < 1.19 \text{ cm}^{-1} $ at the 2 sigma level \cite{Jaeckel:2017tud}.

The expected photon signal can be calculated with \cref{eq:totalFluenceALPVars} for given ALP and SN models, as done in Ref.~\cite{Ferreira:2022xlw}. However, for the SMM observation of SN 1987A we are interested in delay times $ t < t^{\text{max}} = 223 $~s, for a distance to the SN of $ \dSN = 51.4 $~kpc, photon energies $ 25 \text{ MeV} < \omega_\gamma < 100 \text{ MeV} $, and ALP masses $ 10 \text{ keV} < m_a < 350 \text{ MeV} $. Therefore, the condition in \cref{eq:smallAngleTimeCondition} is fulfilled and we can instead use the integrated version of the much simpler \cref{eq:diffFluenceSmallAngle}, speeding up the calculation considerably.

Note that it is possible with the formalism developed in \cref{sec:fluence} to calculate the photon fluence also in the energy range of 10--25~MeV (where SMM also took data) and to determine the time dependence of the signal. However, neither would change the resulting bound as shown in Ref.~\cite{Hoof:2022xbe}.

To calculate the gamma-ray fluence, in addition to the geometric constraints discussed in \cref{sec:fluence}, we also have to implement the constraints
\begin{equation}
    \Theta_{\text{cons.}}(\omega_\gamma, t, \omega_a) \supset \Theta(100 \text{ MeV} - \omega_\gamma) \, \Theta(\omega_\gamma - 25 \text{ MeV}) \, \Theta(223 \text{ s} - t) \, ,
\end{equation}
describing the energy and time window of the SMM observation.
Therefore, using \cref{eq:diffFluenceObsVars} we can calculate the fluence as relevant for the decay bound on ALPs from SN 1987A in the small-angle approximation:
\begin{equation}\label{eq:totalFluenceSmallAngle}
\begin{aligned}
    F_\gamma &= \frac{1}{2\pi \dSN^2} \int_{m_a}^\infty \diff\omega_a \, \int_{\omega_\gamma^{\text{min}}(p_a)}^{\omega_\gamma^{\text{max}}(p_a)} \diff\omega_\gamma \, p_a^{-1} \, \alpSpectrum \, \left( e^{-\frac{R_* m_a}{\tau_a p_a}} - e^{-\frac{t^{\text{max}}}{\tau_a}\frac{2\omega_\gamma}{m_a}} \right) \Theta\left( \Delta\omega_\gamma(p_a) \right) \\[.5cm]
    &= \frac{1}{2\pi \dSN^2} \int_{m_a}^\infty \diff\omega_a \, \left[ \Delta\omega_\gamma(p_a) e^{-\frac{R_* m_a}{\tau_a p_a}} - \frac{\tau_a}{t^{\text{max}}} \frac{m_a}{2} \left( e^{-\frac{t^{\text{max}}}{\tau_a}\frac{2\omega_\gamma^{\text{min}}(p_a)}{m_a}} - e^{-\frac{t^{\text{max}}}{\tau_a}\frac{2\omega_\gamma^{\text{max}}(p_a)}{m_a}} \right) \right] \\
    &\qquad \times p_a^{-1} \, \alpSpectrum \, \Theta\left( \Delta\omega_\gamma(p_a) \right) \, ,
\end{aligned}
\end{equation}
where $ R_* = 3 \cdot 10^{12} \cm $ \cite{Kazanas:2014mca} and the minimal and maximal photon energies for a given ALP energy are:
\begin{equation}
\begin{aligned}
    \omega_\gamma^{\text{min}}(p_a) &= \max \left( 25 \text{ MeV}, \, \frac{1}{2}(\omega_a - p_a), \, \frac{m_a^2 \, R_*}{2 \, p_a \, t^{\text{max}}} \right) \, , \\
    \omega_\gamma^{\text{max}}(p_a) &= \min \left( 100 \text{ MeV}, \, \frac{1}{2}(\omega_a + p_a) \right) \, , \\
    \Delta \omega_\gamma(p_a) &= \omega_\gamma^{\text{max}}(p_a) - \omega_\gamma^{\text{min}}(p_a) \, .
\end{aligned}
\end{equation}
The condition $ \Delta \omega_\gamma(p_a) > 0 $ can be translated into a lower bound on $ p_a $, which does not have a simple analytical form, but can be efficiently evaluated numerically.

\begin{figure}
    \centering
    \includegraphics[width=0.95\textwidth]{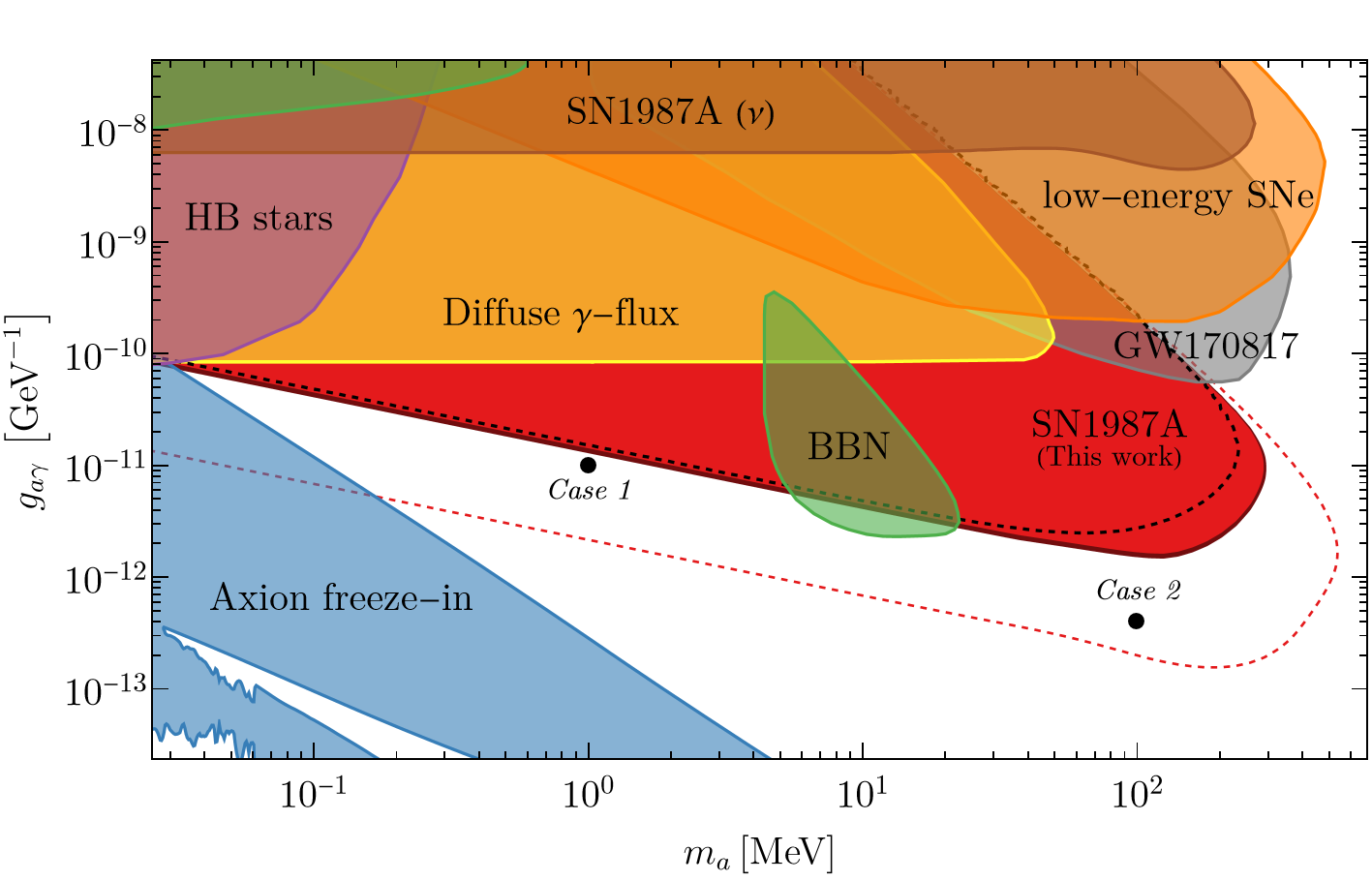}
    \caption{Relevant constraints on the ALP-photon coupling. Here, we have calculated the solid red exclusion regions from gamma-ray decays of ALPs produced in SN 1987A and measured by SMM. Note that the red regions are shown without a boundary, the darker red shade is the $ 2 \sigma $ constraint that is slightly larger than the bright red $ 3 \sigma $ constraint -- see the main text for more details. We have furthermore derived the red dashed contour, which shows an estimate of the sensitivity of \lat~in the event of a future SN (see \cref{sec:lat_data} for details). The black dashed line is the SN 1987A bound found in Ref.~\cite{Hoof:2022xbe}.
    The other, semi-transparent regions are constraints from the energy ALPs could deposit in low-energy SNe (orange) and the diffuse gamma-ray flux due to decays of ALPs produced in all past SNe (yellow), both from Ref.~\cite{Caputo:2022mah}; anomalous cooling during the SN explosion leading to a shorter neutrino burst following SN 1987A (brown) \cite{Lucente:2020whw}; changes in the evolution of horizontal branch stars (purple) \cite{Ayala:2014pea,Carenza:2020zil,Lucente:2022wai}; the non-observation of X-rays after the multi-messenger observation GW170817 of a neutron star merger \cite{Diamond:2023scc}; the irreducible cosmic ALP density from freeze-in production (blue) \cite{Langhoff:2022bij}; as well as from the dissociation of light elements during BBN (green) \cite{Depta:2020wmr}. Note that the BBN bound is the most conservative one presented in Ref.~\cite{Depta:2020wmr}. Depending on the details of the cosmology of the early universe (especially the value of the reheating temperature, here assumed to be 10~MeV), much stronger bounds can be derived, potentially excluding most of the parameter range considered here.
    The two black points mark the ALP parameters of \textbf{case 1 \& 2} discussed in \cref{sec:futureSN}.
    \label{fig:updatedJMRbound}}
\end{figure}
We present the resulting bound on the ALP-parameter space in \cref{fig:updatedJMRbound}. In the red region, too many gamma rays would be produced in conflict with the SMM measurements, where the lighter red corresponds to an exclusion at the 3 sigma level, as done in Ref.~\cite{Jaeckel:2017tud}, while the darker red region is excluded at the 2 sigma level. The other, semi-transparent regions are 
excluded due to constraints found in the literature, see the caption of \cref{fig:updatedJMRbound}.
We find a very good agreement of our bound with the 3 sigma bound of Ref.~\cite{Hoof:2022xbe}, shown as black dashed line, for ALP masses around or below 50~MeV. In that reference, only the Primakoff contribution to the ALP spectrum $ \alpSpectrum $ is taken into account and the spectrum is not calculated directly from a SN model as we do here, but it was rather inferred from the fit to a volume-integrated production spectrum found in Ref.~\cite{Payez:2014xsa}. As the photon-coalescence contribution to the ALP-spectrum, which is the dominant one for masses larger than 70~MeV, is not included in Ref.~\cite{Hoof:2022xbe}, a small discrepancy at larger masses is expected. Thus, as expected, our bound extends to larger masses of about 280~MeV than those found previously reaching only around 200~MeV.\footnote{We thank Sebastian Hoof for making the results of Ref.~\cite{Hoof:2022xbe} available to us for easy comparison.}
Recently, it was pointed out in Ref.~\cite{Diamond:2023scc} that in a small part of the parameter space seemingly excluded by the decay bound, the local density of gamma-ray photons produced by the decaying ALPs outside the SN could in fact be high enough to form a QED plasma through rapid pair-creation $ \gamma \gamma \to e^{+} e^{-} $. This ``fireball'' would radiate energy in the form of x-ray photons of which essentially none would have been detected by SMM such that the bound we derive here does in fact not apply as it is. However, considering the non-observations of x-ray photons in the energy range $ 0.2 - 2 $~MeV by the Pierre Venus Orbiter, also such ALPs that lead to fireball formation are in fact excluded \cite{Diamond:2023scc}. Since the parameter region in question is rather small, lies entirely inside the decay bound, and is also excluded by observations, we do not study this phenomenology further.
Finally, the red dashed line is our estimate of the sensitivity of \lat~to a future nearby SN, which we derive and further discuss in \cref{sec:futureSN}.

\section{Prediction for a gamma-ray spectrum from a future nearby SN}\label{sec:futureSN}
The next Galactic SN explosion will be an important natural laboratory to probe the existence of exotic particles, including ALPs. These events are expected to happen with a rate of $3.2^{+7.3}_{-2.6}$ nearby SNe per century~\cite{Adams:2013ana}, an estimate based on a sparse set of recorded observations during the last millennium. The last nearby SN explosions was SN 1987A, the first event analyzed with modern technologies such as neutrino detectors \cite{Arnett:1989tnf} and the Gamma-Ray Spectrometer onboard the SMM satellite, whose measurements yield one of the strongest bounds on ALPs with masses above 30~keV.
Given the estimated SN rate of around three per century, it is reasonable to expect the next nearby SN event to happen in the near future. In this case, with modern gamma-ray telescopes we expect to observe a signal originating from ALPs with much weaker couplings and/or higher masses than the existing bounds. The \textit{Fermi} Large Area Telescope is sensitive to gamma rays in the range from ~20 MeV to more than 300 GeV, scanning the entire sky every three hours, and it is, therefore, the best available option to study the gamma-ray signal of heavy ALPs from the next nearby SN. As we will show in this section, in the case that no signal would be observed strong constraints can be put on the ALP parameter space. On the other hand, in the favorable case that a signal would indeed be detected, we show which ALP parameters can be determined from this measurement.

We start by describing our data analysis framework in \cref{sec:lat_data}, including a determination of the expected background which yields the sensitivity shown in \cref{fig:updatedJMRbound}; in \cref{sec:FermiFlux} we calculate the expected gamma-ray flux that can be searched for with \lat. Finally, we describe how to fit an observed signal and infer the underlying ALP parameters in \cref{sec:FermiFit}.

\subsection{\lat~simulation and data analysis framework}
\label{sec:lat_data}

To assess the potential of \lat~to constrain the ALP parameter space via a future nearby SN and the subsequent gamma-ray flux due to ALP decay events, we define as our benchmark case a SN with properties similar to SN 1987A. That is:
\begin{itemize} 
\item location in the sky in Galactic coordinates: $\left(\ell, b\right) = (279.70^{\circ}, -31.94^{\circ})$,
\item a distance of $\dSN = 51.4$ kpc and
\item a stellar progenitor of roughly $18\;M_{\odot}$.
\end{itemize}
Since the main effect of an increased distance $ \dSN $ on the gamma-ray signal is a suppression of the number of photons (e.g.~in \cref{eq:diffFluenceSmallAngle} the fluence scales as $ \dSN^{-2} $), these properties are in fact a conservative assumption. The average distance for a SN in the Milky Way is expected to be $ \dSN \sim 10 \kpc $ and the observation of a SN at this distance would improve the sensitivity to $ \gag $ by a factor $ \sim 2.3 $. Note that there are even 31 SN candidates with a distance to the Sun smaller than $ 1 \kpc $, of which Spica is the closest one, at only around 77~pc \cite{Mukhopadhyay:2020ubs}.
The first step in this analysis is to determine the expected gamma-ray background in the direction of a hypothetical SN and this is done in a data-driven way. 

\paragraph{Data selection.} We select about 14.5 years of \lat~data (4th of August 2008 to 19th of January 2023), within a cone of radius $10^{\circ}$ centered on the position of the simulated SN event, satisfying the reconstruction criteria of the event class \texttt{P8R3\_TRANSIENT020\_V3} for \texttt{FRONT+BACK} type events, i.e.~not excluding photons due to their event type. The chosen event class is specifically tailored for transient events such as gamma-ray bursts and timing studies, which generally benefit from enhanced photon statistics and remain relatively unaffected by a higher background fraction and a broader point-spread function.  We consider gamma rays with energies from 25 MeV to 600 GeV, and zenith angles of less than $80^{\circ}$ to reduce the contamination by photons from the Earth's limb. We apply the additional event quality cuts (\texttt{DATA\_QUAL>0 \&\& LAT\_CONFIG==1}). These cuts make sure that no time periods when a particular spacecraft event has affected the quality of the data are taken into account. Moreover, we use a region-of-interest (ROI) cut on the spacecraft file to derive the \lat's Good Time Intervals (GTIs), which singles out time periods when the target was in the field of view of the LAT. All selection, cleaning, manipulation and simulation of \lat~data is conducted via the Fermi Science Tools\footnote{\url{https://github.com/fermi-lat/Fermitools-conda}} (version 2.0.8) \cite{2019ascl.soft05011F}.

\paragraph{Background rate estimation.} The photon signal from ALP decays covers a wide range of time scales depending on the ALP mass. Hence, we expect a varying level of background events that are, by definition, not associated with the SN explosion. To derive the expected number of background events for a specific signal duration $t_{\mathrm{obs}}$ we adopt the statistical approach presented in Ref.~\cite{Meyer:2016wrm} as follows:
\begin{enumerate}
\item We create temporally binned \lat~data for the full dataset using $t_{\mathrm{obs}}$ as bin size.
\item Calculate the LAT exposure $\mathcal{E}$ for each of these bins with the routine \texttt{gtexposure}.
\item Select one temporal bin with non-zero exposure as the ``ON'' region in which the SN ALP-induced gamma-ray signal will be simulated.
\item All other temporal bins are considered ``OFF'' events, $N_{\mathrm{OFF, i}}$, and used to create an estimator for the expected background counts $\hat{b}$ in the ``ON'' region. The value of $\hat{b}$ follows from maximizing the Poisson likelihood function
\begin{equation}
\mathcal{L}\!\left(\left.\vec{N}_{\mathrm{OFF}}\right|b,\vec{\varepsilon}\right) = \prod_i\frac{(\varepsilon_ib)^{N_{\mathrm{OFF, i}}}}{(N_{\mathrm{OFF, i}})!}e^{-\varepsilon_ib} \, ,
\end{equation}
with respect to the background counts $b$,
where $\varepsilon_i = \mathcal{E}_{\mathrm{OFF}, i}/ \mathcal{E}_{\mathrm{ON}}$ is the ratio of exposures in the respective OFF region and ON region. It follows that
\begin{equation}
\hat{b} = \frac{\sum_i N_{\mathrm{OFF, i}}}{\sum_i \frac{\mathcal{E}_{\mathrm{OFF}, i}}{\mathcal{E}_{\mathrm{ON}}}}.
\end{equation}
\end{enumerate}
Since no SN explosion has occurred during the past 14.5 years in this direction, the background estimate does not contain contamination by a potential signal. 

\paragraph{\lat~sensitivity forecast to a future supernova.} Using the data-driven background estimate $\hat{b}$, we can follow Ref.~\cite{Hoof:2022xbe} and \cref{sec:SN 1987A_bound} to forecast the sensitivity of \lat~to a future SN event. We set $t_{\mathrm{obs}} = 24$ h and assume the ON region to start at 297575017 MET (mission elapsed time). We select this particular time period as it roughly features an exposure close to the median exposure for 1 day time slices evaluated over the entire LAT lifetime. The estimator of the average number of background counts amounts to $\hat{b}=677.22$ while the exposure in the ON region reads $\mathcal{E}_{\mathrm{ON}} = 2.62\times10^{7}\;\mathrm{cm}^2\,\mathrm{s}$. Equipped with these numbers we compute the maximal number of photons $s$ expected from ALP decays in the range from 25 MeV to 600 MeV given the ALP mass $m_a$ and as a function of $g_{a\gamma}$. To derive upper bounds on the ALP parameter space we consider the ``ON'' Poisson likelihood function
\begin{equation}
\mathcal{L}\!\left(\left.N_{\mathrm{ON}}\right|s,\hat{b}\right) = \frac{(s + \hat{b})^{N_{\mathrm{ON}}}}{(N_{\mathrm{ON}})!}e^{-(s + \hat{b})} \, ,
\end{equation}
where $ s = \frac{\mathcal{E}_{\rm ON}}{t_{\rm obs}} F_\gamma$ is the expected average signal count, with $ F_\gamma $ calculated by integrating \cref{eq:diffFluenceObsVars} because the small angle approximation does not hold on the full parameter range for the longer delay times considered here, compared to the SMM observation.
We assume for the ON counts $N_{\mathrm{ON}} = \hat{b}$, which is typically done in sensitivity forecasts and states that the measured number of events in the ON region equals the expected background events (the so-called Asimov dataset \cite{2011EPJC...71.1554C}). As counts must be integers, we take the next integer greater than $\hat{b}$. Upper bounds follow from the log-likelihood ratio test and $s\geq 0$ based on
\begin{equation}
\lambda(s) = -2\left[\ln{\mathcal{L}\!\left(\left.N_{\mathrm{ON}}\right|s,\hat{b}\right)} - \ln{\mathcal{L}\!\left(\left.N_{\mathrm{ON}}\right|s=0,\hat{b}\right)}\right]\mathrm{,}
\end{equation}
which solely depends on $g_{a\gamma}$ for fixed $m_a$. The log-likelihood ratio $\lambda$ follows a half-$\chi^{2}$-distribution with one degree of freedom (see Sec.~3.6 of Ref.~\cite{2011EPJC...71.1554C}). In such a scenario the 95\% confidence level (C.L.) upper limit on $s$ is reached when $\lambda$ attains a value of 2.71 (one-sided), i.e.~$\hat{s}=43.75$, which can easily be translated to a constraint on the ALP-photon coupling using \cref{eq:diffFluenceObsVars}. We display the result as a dashed red line in \cref{fig:updatedJMRbound}. As can be seen from there, in the case of a non-observation, \lat~would be able to constrain ALPs with a photon coupling more than an order of magnitude below the current limit from SN 1987A and with masses more than a factor of 2 larger. This is an interesting result for our forecast, even considering that we conservatively assume a large distance from the SN (around 50~kpc).

\subsection{Predicted photon flux}
\label{sec:FermiFlux}

When we have access to the time and energy dependence of an observed photon signal, the quantity we have to compare to those measurements is not the total fluence, but rather the spectral flux of gamma rays
\begin{equation}
     \photonSpectralFlux \equiv \frac{\mathrm{d}^2 F_\gamma}{\mathrm{d}\omega_\gamma \mathrm{d}t} \, .
\end{equation}
Following \cref{sec:fluence}, we can calculate this spectral flux by integrating the respective differential fluence formulas over the observation angle in general (\cref{eq:diffFluenceObsVars}), or the ALP energy in the small-angle approximation (\cref{eq:diffFluenceSmallAngle}). 

We verified that even if the small-angle condition is violated, the angular spread of the signal is at most $\sim\mathcal{O}(1^{\circ})$, in agreement with Ref.~\cite{Jaeckel:2017tud}.
The angular resolution of \lat, defined as the $68\%$ containment radius of the LAT's point spread function\footnote{See \url{https://www.slac.stanford.edu/exp/glast/groups/canda/lat_Performance.htm} for details.}, depends on the chosen LAT event class and type. In the considered case and around 25 MeV it is $\mathcal{O}(10^{\circ})$. Thus the gamma-ray burst will appear as a point source. Moreover, this large angular spread is achieved in regions of the parameter space in which neither SMM nor \lat~have any sensitivity because the photons' arrival time is distributed over a time-span of order 10 years, such that the fluxes at any point in time are very small.

For observation times, angles, and energies, in which \cref{eq:smallAngleTimeCondition} is always fulfilled, we can use the differential fluence in \cref{eq:diffFluenceSmallAngle}. In the small-angle approximation, the resulting flux as observable by \lat, would therefore be
\begin{equation} \label{eq:smallAngleFlux}
    \photonSpectralFlux = \frac{1}{\pi \dSN^2} \frac{\omega_\gamma}{\tau_a \, m_a} \, e^{-\frac{t}{\tau_a} \frac{2\omega_\gamma}{m_a}} \int_{\omega_\gamma + \frac{m_a^2}{4\omega_\gamma}}^{\infty} \mathrm{d}\omega_a \, p_a^{-1} \, \alpSpectrum \Theta\left(p_a - \frac{m_a^2}{2\omega_\gamma t} R_\star\right)
    \, .
\end{equation}
Note that for $ m_a \ll \omega_\gamma $ this expression agrees with Refs.~\cite{Caputo:2021rux,Oberauer:1993yr,Jaffe:1995sw}. 
The flux falls off exponentially with the delay time $ t $ on a time-scale
\begin{equation}
    \frac{m_a}{2 \omega_\gamma} \tau_a = 76.6 \text{ days } \left(\frac{\gag}{10^{-11} \text{ GeV}}\right)^{-2} \left(\frac{m_a}{1 \text{ MeV}}\right)^{-2} \left(\frac{\omega_\gamma}{100 \text{ MeV}}\right)^{-1} \, .
\end{equation}
On the other hand, for small $ t $ and a fixed photon energy, the Heaviside function in \cref{eq:smallAngleFlux} will suppress the flux since the ALP spectrum decreases exponentially at large $ \omega_a $. When both of these cut-offs can be ignored, the flux is essentially constant in time.

In what follows, we show a simple analytical formula for the gamma-ray flux that is helpful to efficiently fit observations from a future nearby SN. In fact, \cref{eq:smallAngleFlux} is simple enough to be integrated analytically if we assume an analytical form for the ALP spectrum as~\cite{Payez:2014xsa,Calore:2021hhn}
\begin{equation} \label{eq:alpSpectrumFit}
    \alpSpectrum \simeq \gag^2 \, C_0 \left( \frac{\omega_a}{\omega_a^0} \right)^\alpha \exp\left[-(1+\alpha) \frac{\omega_a}{\omega_a^0}\right] \, ,
\end{equation}
where $ \omega_a^0 $ is related to the average ALP energy in the case of light ALPs, $ \alpha $ is a dimensionless spectral index, and $ C_0 $ is a normalization constant that is determined by properties of the SN, such as its temperature, density and time during which ALPs are effectively produced.
Furthermore, in order to obtain an analytical result for the spectral photon flux, we have to assume that ALPs are ultrarelativistic, such that $ p_a \simeq \omega_a $ is a good approximation for the integral in \cref{eq:smallAngleFlux}. Since the peak of the ALP spectrum is around $ \omega_a \sim 100 $~MeV for light ALPs, the highest ALP mass for which we obtain a simple analytical result is $ m_a \sim 10 $~MeV.

Using the analytical form of the spectrum for ultrarelativistic ALPs, \cref{eq:smallAngleFlux,eq:alpSpectrumFit} yield:
\begin{equation}
\begin{aligned} \label{eq:alphaFitURFlux}
    \frac{\mathrm{d} \Phi_\gamma}{\mathrm{d} \omega_\gamma}\Big\rvert_{m_a < 10 \text{ MeV}}
    &= A \, \omega_\gamma\, e^{- B \, t \, \omega_\gamma} \, \Gamma\left[\alpha, (1+\alpha) \frac{\omega_a^{\text{min}}}{\omega_a^0} \right] \, ,
\end{aligned}
\end{equation}
where $ \Gamma(s,x) $ is the incomplete gamma function, and we have defined the parameters
\begin{equation}
\begin{split}
\label{eq:FitParam}
    A = \frac{C_0 (1 + \alpha)^{-\alpha}}{64 \pi^2 \, \dSN^2} \, \gag^4 m_a^2 \, , \qquad
    B = \frac{1}{32\pi} \, \gag^2 m_a^2 \, ,\\
    \omega_a^{\text{min}}(t, \, \omega_\gamma) = \max \left( \omega_\gamma + \frac{m_a^2}{4 \omega_\gamma} , \, m_a \sqrt{1 + \left(\frac{m_a \, R_*}{2 \, t \, \omega_\gamma}\right)^2} \right) \, .
\end{split}
\end{equation}
We will use this approximated, analytical form of the flux as a model to fit an ALP signal as it would be observed by \lat~in the following section.

\subsection{Fitting the spectrum: What can we learn from a potential \lat~observation?}
\label{sec:FermiFit}
In \cref{subsec:analysis} we present our simulation and analysis pipeline developed to reconstruct the ALP parameters by examining the gamma-ray signal associated with a potential nearby SN event as observed by the \textit{Fermi}-LAT instrument. The results of the fitting procedure are discussed in \cref{subsec:fitResults}.
\subsubsection{Gamma-ray signal simulation and analysis}
\label{subsec:analysis}
\paragraph{Fitting model and parameters.}
Assuming that the \lat~instrument will measure a gamma-ray signal from decaying ALPs with $ m_a \lesssim 10 $~MeV produced in a potential nearby SN, as discussed in the previous section it should be possible to fit the flux with the function
and parameters defined in \cref{eq:alphaFitURFlux,eq:FitParam}.
To put it the other way around: if a detected gamma-ray flux can be well fitted by \cref{eq:alphaFitURFlux}, this is an indication that it could originate from the decay of light ALPs.

If the exponential decay with $ t \cdot \omega_\gamma $ is observable (this requires relatively late observation times, since large photon energies lead to a fast decay due to the gamma function, overlapping with the exponential factor), we can immediately determine $ \gag \cdot m_a \sim \sqrt{B} $. If the exponential decay is not visible in the data, the non-analyticities stemming from the maximum function in \cref{eq:alphaFitURFlux} could determine $ m_a $ and $ R_* $, while in that case the degeneracy of $ C_0 $ and $ \gag^2 $ makes the exact determination of the coupling only from the measured gamma-ray spectra impossible. Finally, if also no kinks in the spectra are discernible only $ \alpha $ and $ \omega_a^0 $, properties of the ALP spectrum and therefore of the SN, can be inferred from the fit alone.
The only known quantity involved in the factor $A$ is the SN distance, while much larger uncertainties are associated with models predicting the ALP spectral parameters. They can, however, still be estimated: assuming that ALPs are produced by Primakoff conversion only (an assumption valid for $m_{a} < 70$~MeV)  Ref.~\cite{Calore:2021hhn} found the following spectral parameters
	\begin{equation}
    \label{eq:SN_ALP_params}
	\begin{split}
	    \frac{C_0(M)}{10^{78} \text{ MeV}} &= (1.73\pm 0.172) \frac{M}{M_{\odot}} - 9.74 \pm 2.92 \, ,\\
	 	\frac{\omega_a^0(M)}{\text{MeV}} &= (1.77\pm 0.156)  \frac{M}{M_{\odot}} + 59.3 \pm 2.65  \, ,\\
		\alpha (M) &= (-0.0254 \pm 0.00587) \frac{M}{M_{\odot}} + 2.94 \pm 0.0997 \, ,\\
	\end{split}
	\end{equation}
as function of the SN progenitor mass $ M $, where the quoted errors are the $ 1 \sigma $ uncertainties. Note that our $ C_0 $ is not identical to the quantity $ C $ defined in Ref.~\cite{Calore:2021hhn}.
Using these estimates, one can infer the value of the combination $ \gag^4 m_a^2 $ from the best-fit value of $A$, even if the spectral features discussed above are not observable in the data, as we will discuss in the rest of this section.

\begin{figure}
\vspace{0.cm}
\begin{center}
\includegraphics[width=0.49\columnwidth]{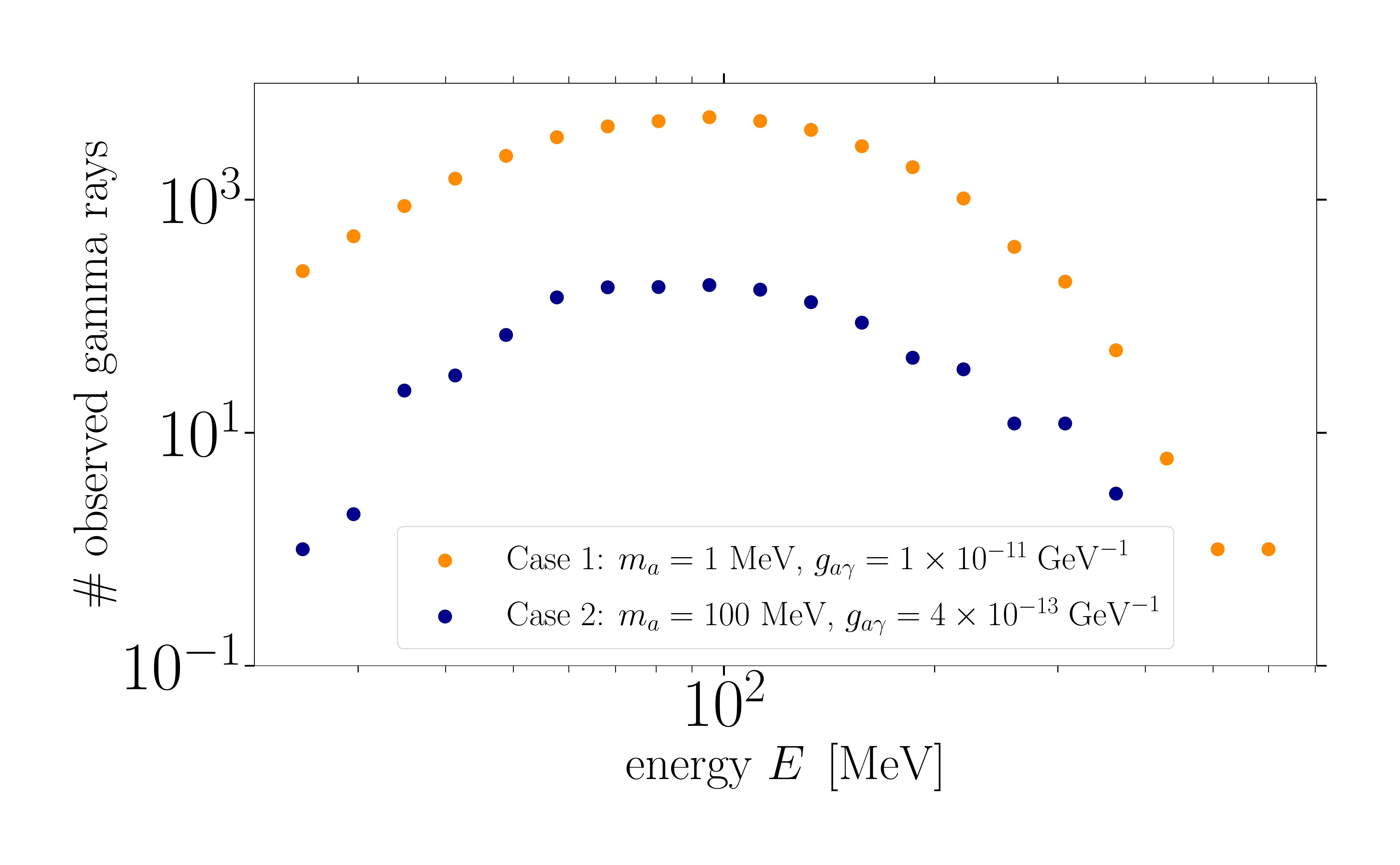}
\hfill
\includegraphics[width=0.49\columnwidth]{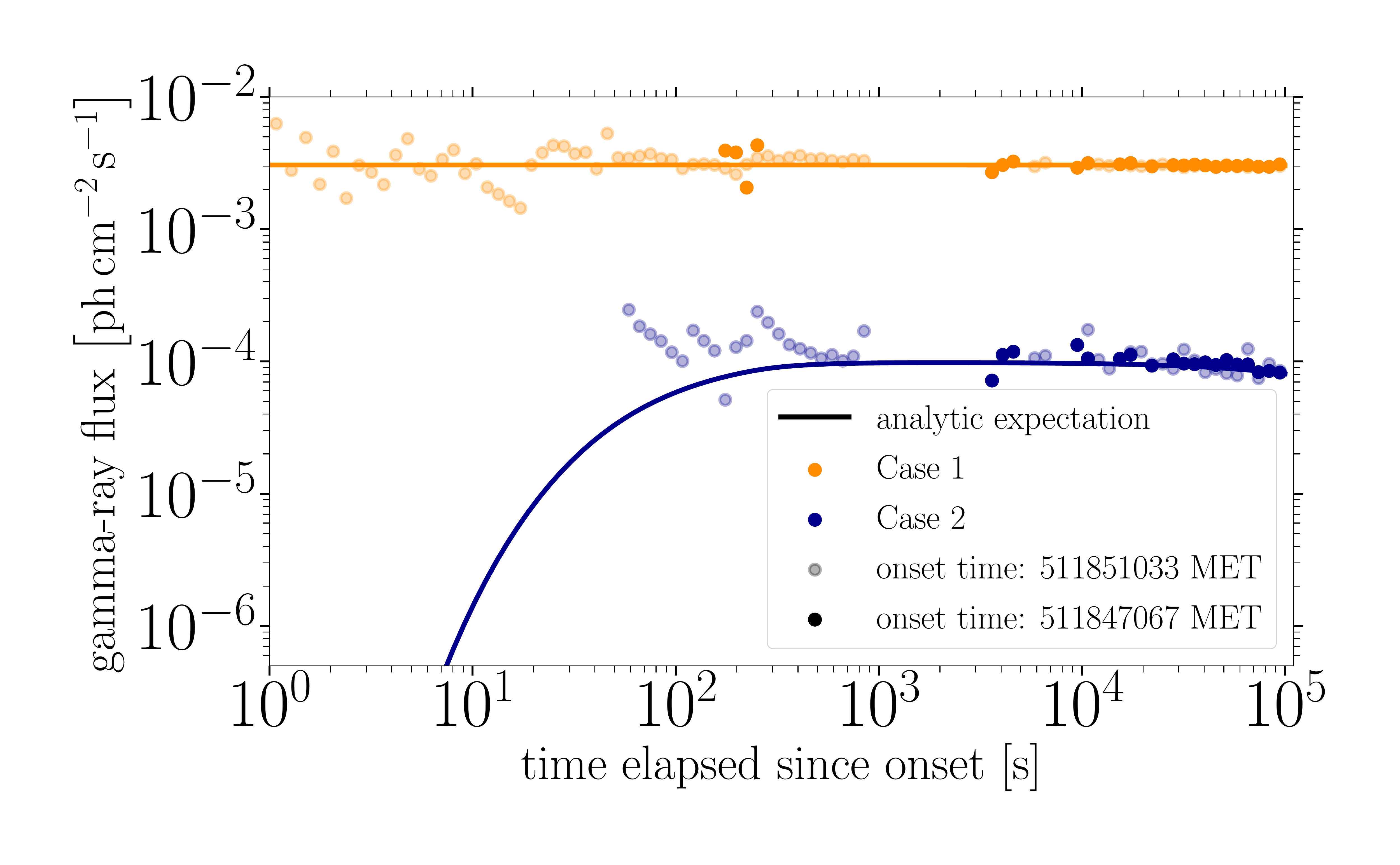}
\caption{(\textit{Left}:) Simulated mock-observations of time-integrated gamma-ray spectra of a future SN induced by ALP decays. The spectra reflect the number of detected photons over a total observation time of $t_{\mathrm{obs}} = 10^5$ s in the energy range from 25 to 600 MeV binned into 20 logarithmically spaced bins. We consider two different scenarios: \textbf{(case 1}) $g_{a\gamma} = 1\times10^{-11}$ GeV$^{-1}$, $m_a = 1$ MeV (orange data points); (\textbf{case 2}) $g_{a\gamma} = 4\times10^{-13}$ GeV$^{-1}$, $m_a = 100$ MeV (blue data points). (\textit{Right}:) Simulated light curves of the same ALP-induced gamma-ray signals integrated over the energy range from 25 to 600 MeV. The analytic expectations for the light curves of both cases are shown as solid lines adhering to the color style of the left figure. The opaque data points reflect our benchmark case with SN onset time of $t_{\mathrm{SN}} = 511847067.0$ MET while the transparent points exemplify a situation where the first seconds of the SN are detected by the \lat~($t_{\mathrm{SN}} = 51185103.0$ MET). We plot the light curves as observed gamma-ray flux, i.e.~detected counts per temporal bin divided by the associated exposure.
}
\label{fig:ALP_spectra}
\end{center}
\end{figure}

\paragraph{General setup.} To assess how much we can learn about ALPs with a realistic observation of a future nearby SN, we resort to the \emph{Fermi}~Science Tools and simulate two examples of observed mock-signals based on the gamma-ray fluence predicted by our calculations. We assume that data are taken for $t_{\mathrm{obs}} = 10^5 $~s (roughly a day) in the range from 25 MeV to 600 MeV using the same SN position and \lat~data selection criteria listed in \cref{sec:lat_data}. For definiteness, we set $t_{\mathrm{SN}} = 511847017.0$ MET as the simulation starting time. Following the reasoning in \cref{sec:lat_data}, also this onset time guarantees an exposure close to the observed median for 1 day periods. In contrast to the value adopted in \cref{sec:lat_data} (which is located around the start of the LAT's mission), this choice of $t_{\mathrm{SN}}$ is at the later end of the LAT's mission elapsed time (22nd of March 2017).
The two test cases of ALP parameters for which we generate mock observations are: 
\begin{itemize}
    \item \textbf{Case 1}: $g_{a\gamma} = 1\times10^{-11}$~GeV$^{-1}$, $m_a = 1$~MeV, a relatively light ALP with a coupling close to the maximal allowed value
    \item \textbf{Case 2}: $g_{a\gamma} = 4\times10^{-13}$~GeV$^{-1}$, $m_a = 100$~MeV, a heavier ALP with a coupling a factor $\sim5$ below the maximal allowed value.
\end{itemize}
We display the position of these values in the unexplored ALP parameter space in \cref{fig:updatedJMRbound}. From the complete, numerical description in \cref{eq:smallAngleFlux} the mock photon counts $\bm{S_0} = \mathcal{E} \cdot \Phi_\gamma $ are simulated using the \textit{Fermi} Science Tools  with the routine \texttt{gtobssim} based on the \texttt{FileSpectrum} class\footnote{See \url{https://fermi.gsfc.nasa.gov/ssc/data/analysis/scitools/obssim_tutorial.html} for a more detailed description of how to work with this \emph{Fermi} Science Tools routine.} and binned in 20 logarithmically spaced energy bins over the full energy range, and $101$ logarithmically spaced temporal bins from $ 0.5 $ s to $ 10^5 $ s after $t_{\mathrm{SN}}$. In \cref{fig:ALP_spectra} we show two representations of the ALP-induced mock signals for both \textbf{cases 1 \& 2} in orange and blue, respectively: The left panel contains the time-integrated gamma-ray spectrum while the right panel displays the light curve of the signal integrated over the full energy range.
The visible scatter of the light curves is due to the Poisson nature of the gamma-ray detection and the rather short time periods, $\mathcal{O}(\leq10\,\mathrm{s})$, in the first few temporal bins (logarithmic spacing).
Our benchmark choice for the initial time of observation $ t_{\rm SN} $ is shown in opaque colors. From the light curves we see that for this choice only the later part of the signal is observed.
In contrast, the transparent data points demonstrate the more fortunate case where the full signal would be observed. The time dependence of the flux may contain valuable information about the ALP parameters that could even break the degeneracy between $\gag$ and $m_a$ (see the discussion in the beginning of \cref{sec:lat_data}).
In our two cases, the most visible difference in the time dependence would only be a later onset of the observed signal for {\bf case 2} -- as is expected due to the time dependence of $ \omega_a^{\rm min} $ until up to $ \mathcal{O}(100) $~s for heavy ALPs (see \cref{eq:alphaFitURFlux}). However, the potential to observe this initial phase strongly depends on the pointing position of \lat~at the moment of the SN explosion (as indicated by the differences between the opaque and transparent light curves in \cref{fig:ALP_spectra}) and the coupling $ \gag $; the latter due to a lack of emitted photons in the early phase. 
With our benchmark case, we make the conservative assumption that this information is not available.

Given the mock observations, we fit the signals using the simplified spectrum in \cref{eq:alphaFitURFlux}, in order to have an analytical expression for the flux.\footnote{In principle, one could also use the numerical expression in \cref{eq:smallAngleFlux} to fit the observation. However, this is computationally costly and will not lead to a significantly different result in most of the parameter space to which~\lat is sensitive for $ m_a \lesssim 10 $~MeV since \cref{eq:alphaFitURFlux} is a good approximation here.} This is done analogous to the mock observations with the \textit{Fermi} Science Tools, where we use a linear interpolation of the spectra generated at the simulated parameter grid nodes as the final signal model $\bm{S}$.
For our fitting model, we assume $ B = 0 $ and $ \omega_a^{\text{min}} = \omega_\gamma $, i.e.~a time-independent signal, due to the essentially flat light curves in \cref{fig:ALP_spectra} for our benchmark case. This leaves us with three parameters ($A$,  $\omega_a^0$, $\alpha$) determining the resulting gamma-ray spectrum. In \cref{tab:spectrum_priors} we list the prior ranges for these parameters, which we employ to simulate model data for our analysis. In practice, we prepare a regular grid of parameter combinations following the stated prior ranges. Each of these combinations uniquely determines a gamma-ray spectrum for which we simulate 30 different Poisson realizations in order to derive a mean expectation for our ALP-induced gamma-ray model.

Note that the second ALP parameter scenario ($m_a = 100$ MeV) is outside of the range of validity of the simplified formula for the flux in \cref{eq:alphaFitURFlux} and of our assumption that the flux is time-independent. Hence, we expect a lower goodness-of-fit and potentially values for the spectral parameters that are incompatible with the form of the SN ALP spectra. We will use this case to demonstrate in \cref{subsec:fitResults} that in the event of an observation, one can at least decide whether the ALP mass is above or below $ \sim 10 ~\MeV $, even though only the product $ \gag^2 m_a \sim \sqrt{(1 + \alpha)^\alpha A} $ is determined directly by the fit.

\begin{table*}
    \centering
    \begin{tabular}{l c c c c}
         \hline
         ALP scenario & $\pi(A)$ & $\pi(\omega_a^0)$ & $\pi(\alpha)$ & \\
         \hline
         \hline
Case 1 & $\left[5, 15\right]\times10^{-7}$ & $\left[85, 90\right]$ & $\left[2.2, 2.4\right]$ \\
Case 2 
& $\left[0.5, 5\right]\times10^{-12}$ & $\left[100, 115\right]$ & $\left[7.95, 8.65\right]$ \\
         \hline
    \end{tabular}
    \caption{Summary of the ALP scenarios considered in this work as well as the assumed parameter prior ranges $\pi(\cdot)$ for the simulation of the spectral gamma-ray model in \cref{eq:alphaFitURFlux}. The units of parameter $A$ are $\left[\mathrm{cm}^{-2}\mathrm{s}^{-1}\mathrm{MeV}^{-2}\right]$, $\omega_a^0$ 
    is given in units of $\left[\mathrm{MeV}\right]$, and $ \alpha $ is dimensionless. \label{tab:spectrum_priors}}
\end{table*}

\paragraph{Spectral parameter inference.} Our goal is to infer information about the nature of the ALPs that can cause the observed gamma-ray emission from a future nearby SN. To this end, we fit our model in \cref{eq:alphaFitURFlux} using the simulated spectra to the mock observations. Thereby we reconstruct the parameters of the initial model, which are related to the fundamental ALP parameters according to \cref{eq:FitParam}. Quantitatively, we employ a generalized Poisson likelihood function:
\begin{equation}
\label{eq:poisson_likelihood}
\begin{split}
\mathcal{L\!}\left(\left.\bm{\mu} = \bm{S} + \bm{B} + \bm{\delta B}\right|\bm{n}\right) =& \prod_{i=1}^{N_E}  \frac{\left(S_{i} + B_i + \delta B_i\right)^{n_{i}}}{\left(n_{i}\right)!}e^{-\left(S_{i} + B_i + \delta B_i\right)} \times \\ 
  &\times\exp{\left[-\frac{1}{2}\sum_{j,k=1}^{N_E} \delta B_j\,\left(K^{-1}\right)_{jk}\,\delta B_k\right]}\mathrm{,}
\end{split}
\end{equation}
where $\bm{S}$ denotes the ALP-induced gamma-ray model describing the signal (averaged over multiple Poisson realizations),  $\bm{B}$ refers to the expected background and $\bm{\delta B}$ are the fluctuations (positive or negative) per energy bin induced by the finite energy resolution of the instrument, whereas $\bm{n}$ refers to the observed mock data (single Poisson realization). The index $i$ labels different energy bins ($N_E = 20$ in total) of the data or model gamma-ray spectra. 

\begin{figure}[t!]
\vspace{0.cm}
\begin{center}
\includegraphics[width=0.40\columnwidth]{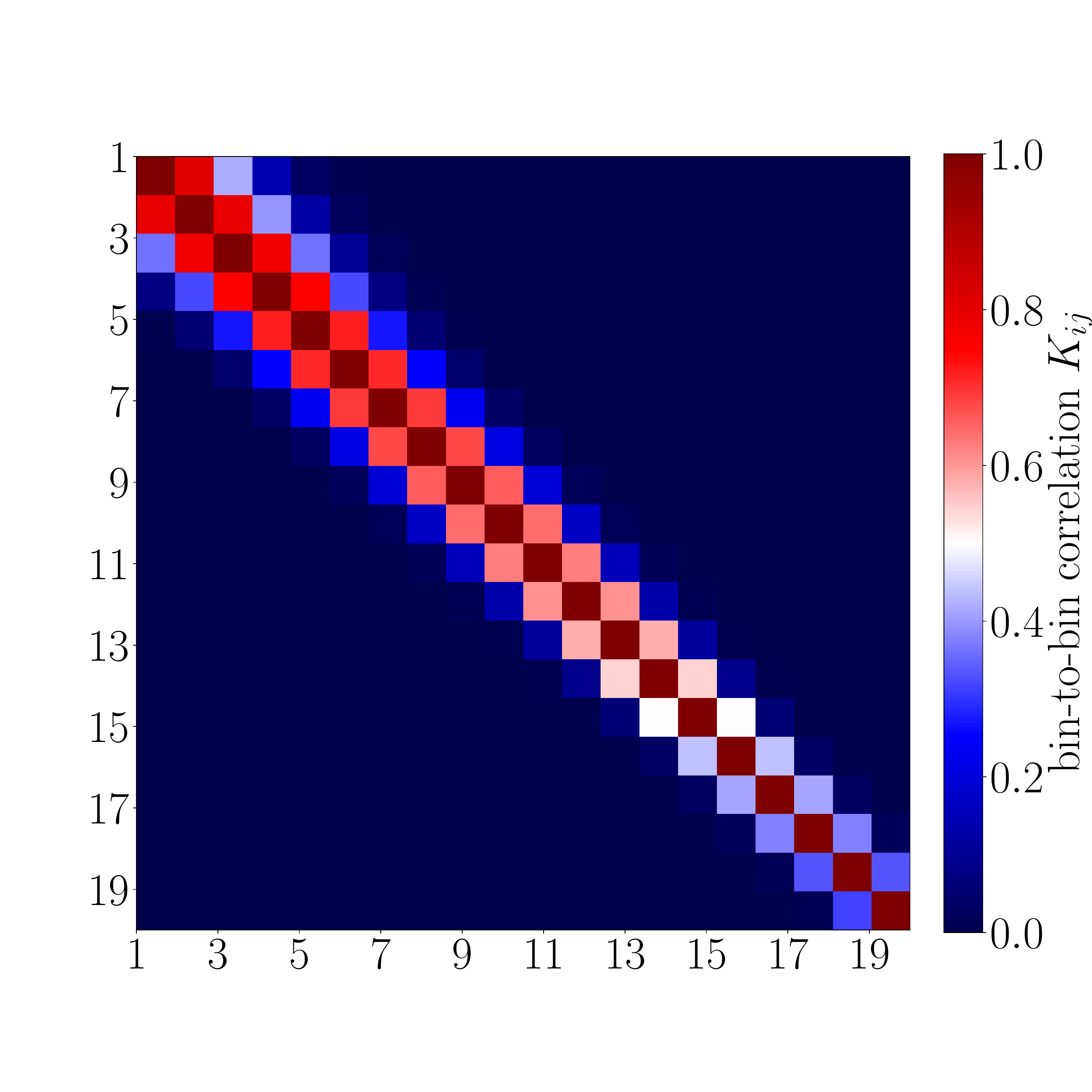}
\caption{Covariance matrix of spectral fluctuations $\bm{\delta B}$ for the \texttt{P8R3\_TRANSIENT020} event class and \texttt{FRONT+BACK} converted events. The spectra are binned from 60 to 600 MeV into 20 logarithmically spaced energy bins represented in the matrix. The correlation follows the parametric description of \cref{eq:energy_cov} with $\sigma(E)\equiv 1$.}
\label{fig:cov_matrix}
\end{center}
\end{figure}

Lastly, the covariance matrix $K_{ij}$ parameterizes the correlation between the spectral fluctuations $\bm{\delta B}$ that is dictated by the energy dispersion of the LAT for the selected event class and event type. The values are taken from the respective LAT instrument response function file provided via the \emph{Fermi} Science Tools. Quantitatively, the covariance matrix is defined as:
\begin{equation}
\label{eq:energy_cov}
K\!\left(E_i, E_j\right) = \sigma\!\left(E_i\right)\sigma\!\left(E_j\right)\exp{\!\left[-\frac{1}{2}\left(\frac{\ln{E_i/E_j}}{w_i}\right)^2\right]}\mathrm{,}
\end{equation}
where $\sigma(E_i)$ describes the percentage fluctuation of the signal at energy $E_i$ while $w_i$ is the energy correlation length at this particular energy according to the energy dispersion. For the scale $\sigma(E)$ we adopt a constant value of 15$\%$, which is compatible with the scatter of spectra that we obtain from simulations. The covariance matrix is displayed in \cref{fig:cov_matrix} (setting $\sigma(E)\equiv 1$).

In practice, we derive the estimator of background counts $\hat{\bm{B}}$ in the simulated time interval analogously to the case considered in \cref{sec:lat_data} but accounting for the energy binning. The observed mock data is consequently given by drawing a Poisson realization from $\hat{\bm{B}}$ and setting: $\bm{n} = \bm{S_0} + \bm{\hat{B}_0}$. We aim at computing the posterior distributions $\bm{\pi}(\left.\cdot\right|\bm{n})$ for the model parameters, i.e.~the parameters of \cref{eq:alphaFitURFlux}, in a Bayesian approach. To this end, we scan the logarithm of the likelihood function in \cref{eq:poisson_likelihood} in the following way:
\begin{enumerate}
    \item The fluctuations $\bm{\delta B}$ are treated as nuisance parameters. In the first step, we profile over these nuisance parameters in a maximum likelihood fit with \texttt{iminuit}~\cite{iminuit}, i.e.~for fixed values of $(A, \omega_a^0, \alpha, \left[m_a\right])$ the best-fitting values for $\bm{\delta B}$ are derived by minimizing $-2\ln{\mathcal{L}}$, which defines the profiled log-likelihood function $-2\ln{\mathcal{L}}_{\mathrm{prof}}$.
    \item $-2\ln{\mathcal{L}}_{\mathrm{prof}}$ is now employed to derive the  (marginal) posterior distributions for the remaining model parameters using \texttt{MultiNest}~\cite{Feroz:2008xx} specifying 1000 live points and an evidence tolerance of 0.2. 
\end{enumerate}

\subsubsection{Results and interpretation} \label{subsec:fitResults}
\begin{figure}
    \centering
    \includegraphics[width=.9\textwidth]{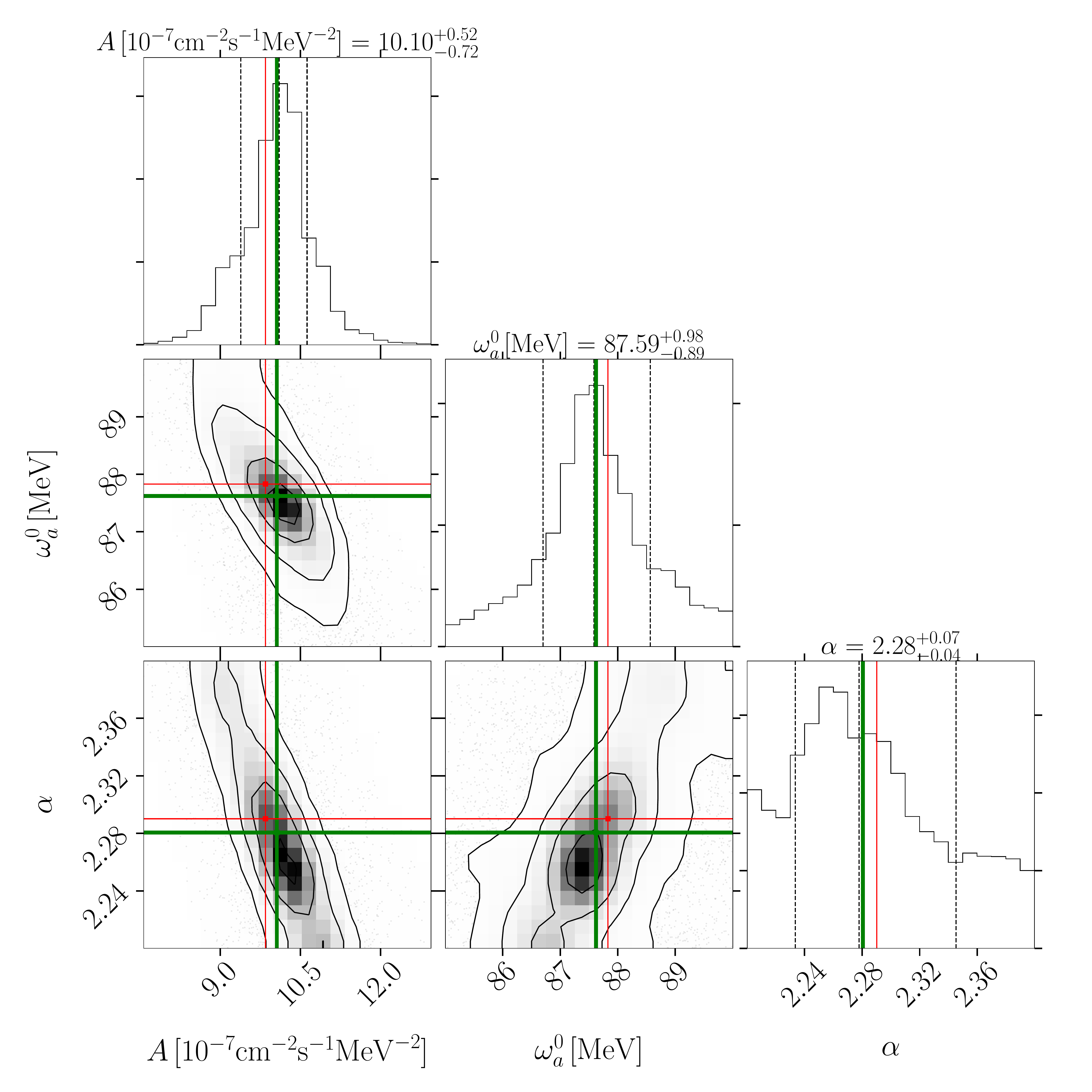}
    \caption{Best-fitting values and posterior distributions of the reconstructed ALP spectral parameters $A, \omega_a^0$ and $\alpha$ for an ALP-induced gamma-ray signal of a future SN characterized by $m_a = 1$ MeV, $g_{a\gamma} = 10^{-11}$ GeV$^{-1}$ (\textbf{case 1}). We overlay the marginal two-dimensional  posterior distributions with the best-fitting parameter values using the analytic formula in \cref{eq:alphaFitURFlux} (before the simulation with the Fermi Science Tools) while the green values are denoting the parameter values maximizing \cref{eq:poisson_likelihood}. The marginal one-dimensional posterior distributions for each parameter show the $16\%$, $50\%$ (median) and $84\%$ quantiles as black dashed lines, whose numerical values are also stated in the title of each marginal posterior.}
    \label{fig:fit_alpParams1}
\end{figure}

\paragraph{Case 1.} This scenario features three effectively accessible parameters, $A, \omega_a^0$ and $\alpha$. The parameter $B$ can safely be neglected, and $ \omega_a^{\text{min}} \simeq \omega_\gamma $ is a good approximation. Assuming uniform priors for all three fit parameters, we obtain the marginal posterior distributions shown in \cref{fig:fit_alpParams1}.
The obtained posteriors strikingly reveal \lat's potential concerning the search for ALPs in the MeV range. The parameters are all well reconstructed with similar relative uncertainties of $\sim 5\% $.
The largest source of uncertainty on the parameters is most likely the LAT's sizeable energy dispersion in the energy range from 25 MeV to 600 MeV, which smears the exact spectral shape. As argued in \cref{sec:FermiFlux}, the simplified fit function for the ALP-induced gamma-ray flux is overall a good description for $ m_a \lesssim 10 $~MeV. We indicate the parameters found by fitting \cref{eq:alphaFitURFlux} directly to the numerically calculated spectrum (determined by \cref{eq:smallAngleFlux}) as red points in the marginal posteriors. They are in good agreement with the parameter values maximizing the generalized Poisson likelihood in \cref{eq:poisson_likelihood}, which underlines the quality of the fit function for this case.

\begin{figure}
    \centering
    \includegraphics[width=.60\textwidth]{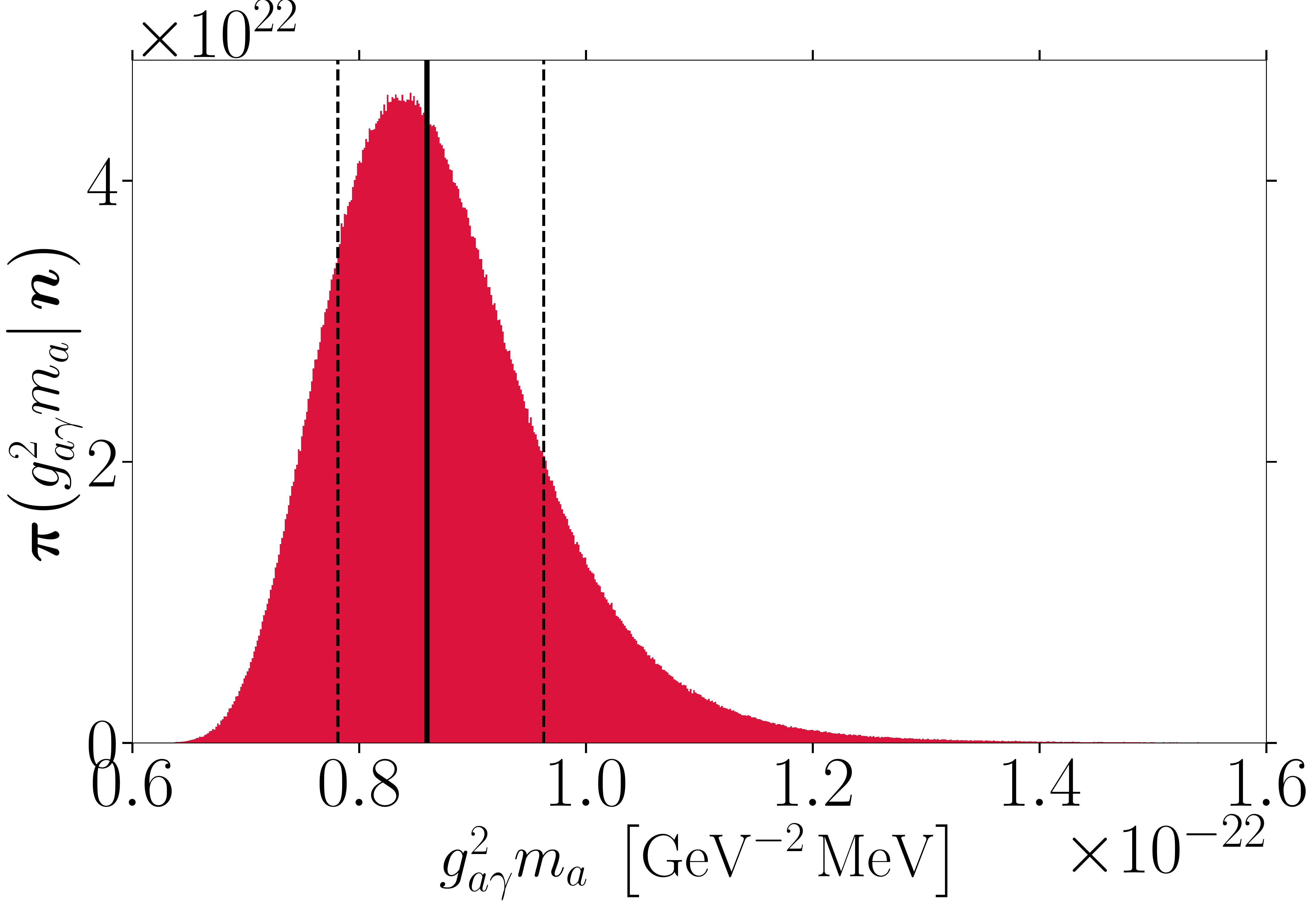}\\
    \includegraphics[width=.49\textwidth]{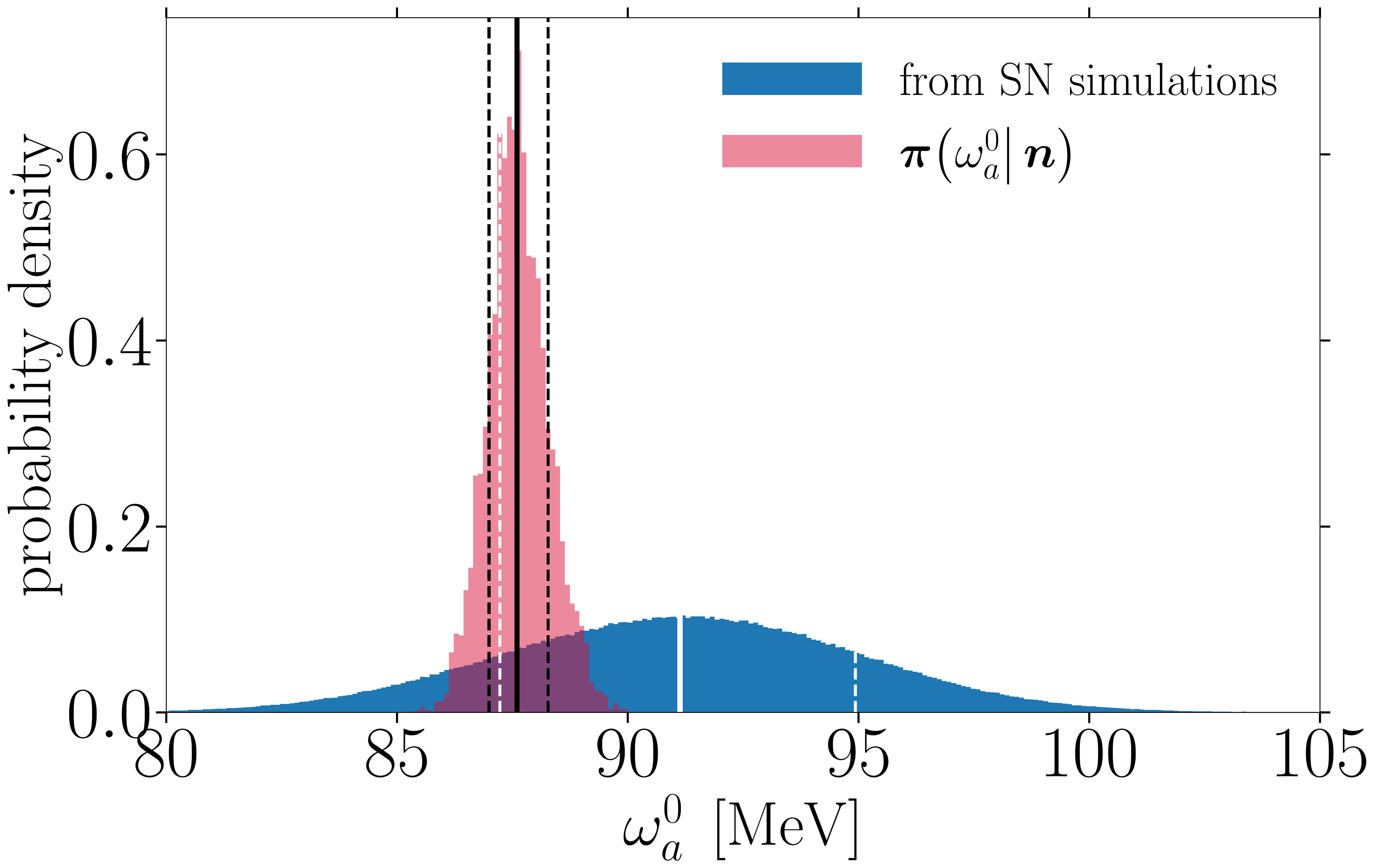}
    \includegraphics[width=.49\textwidth]{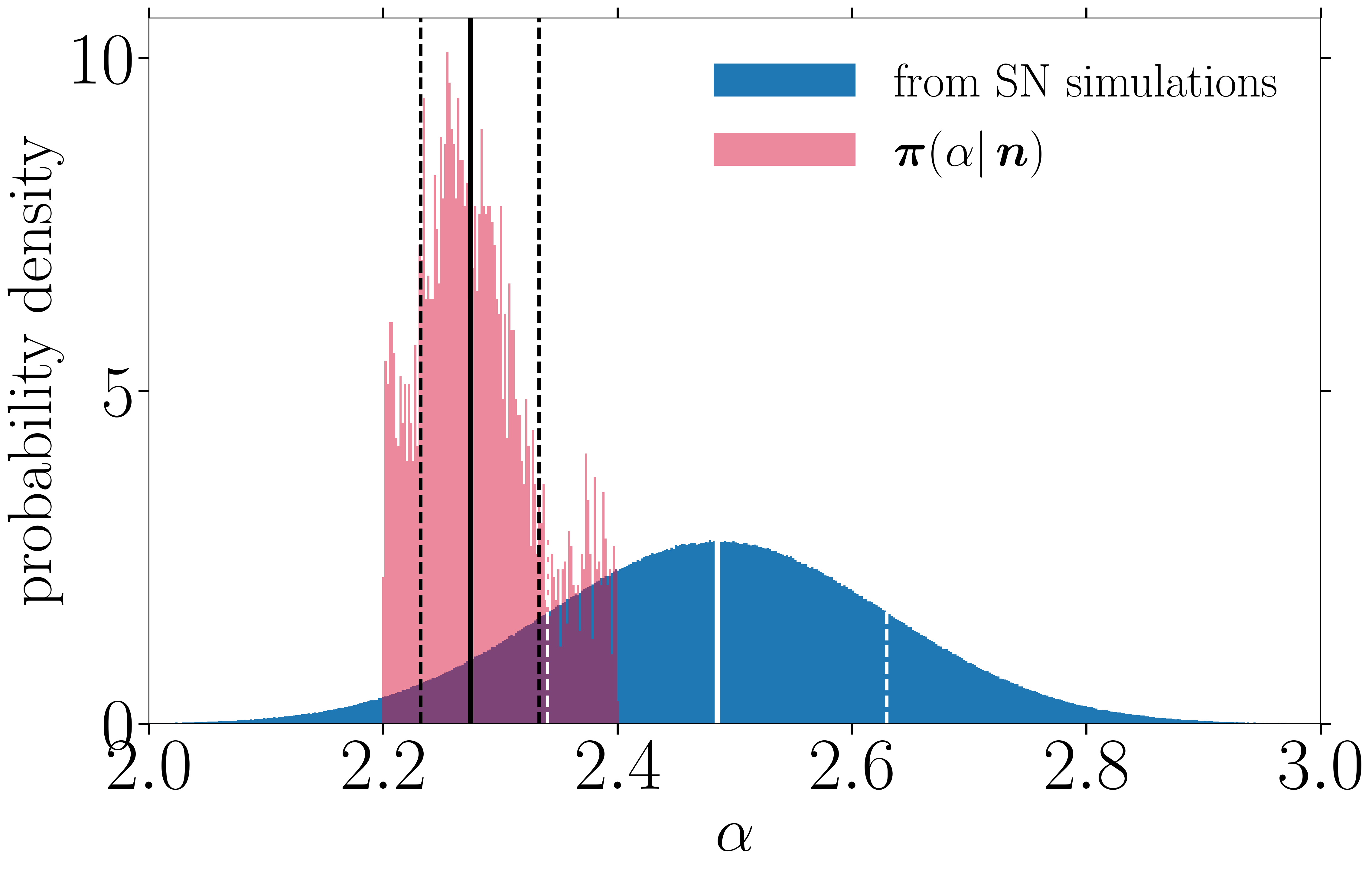}
    \caption{Marginal one-dimensional posteriors (red) for various parameters describing the ALP-induced gamma-ray signal accessible in \textbf{case 1}. (\emph{Left}:) $\bm{\pi}\!\left(\left.g_{a\gamma}^2 m_a\right|\bm{n}\right)$ sampled from the joint posterior distribution derived via \texttt{MultiNest} and the distribution of $C_0$ according to \cref{eq:SN_ALP_params} for an $18\;M_{\odot}$ stellar progenitor. The black, solid vertical line marks the median of the posterior sample while the black, dashed lines denote the $16\%$ and $84\%$ quantiles, respectively. (\emph{Middle}:) Comparison of the observationally inferred posterior distribution $\bm{\pi}\!\left(\left.\omega_a^0\right|\bm{n}\right)$ compared to the expected scatter (blue) of the same quantity according to numerical simulations of ALP production in SNe (see \cref{eq:SN_ALP_params}). The white vertical lines represent the $16\%$, $50\%$ (median) and $84\%$ quantiles of the blue sample. (\emph{Right}:) Same as the middle panel but for $\bm{\pi}\!\left(\left.\alpha\right|\bm{n}\right)$.}
    \label{fig:ALP_posteriors_ALPparams1}
\end{figure}

\begin{figure}
    \centering
    \includegraphics[width=.9\textwidth]{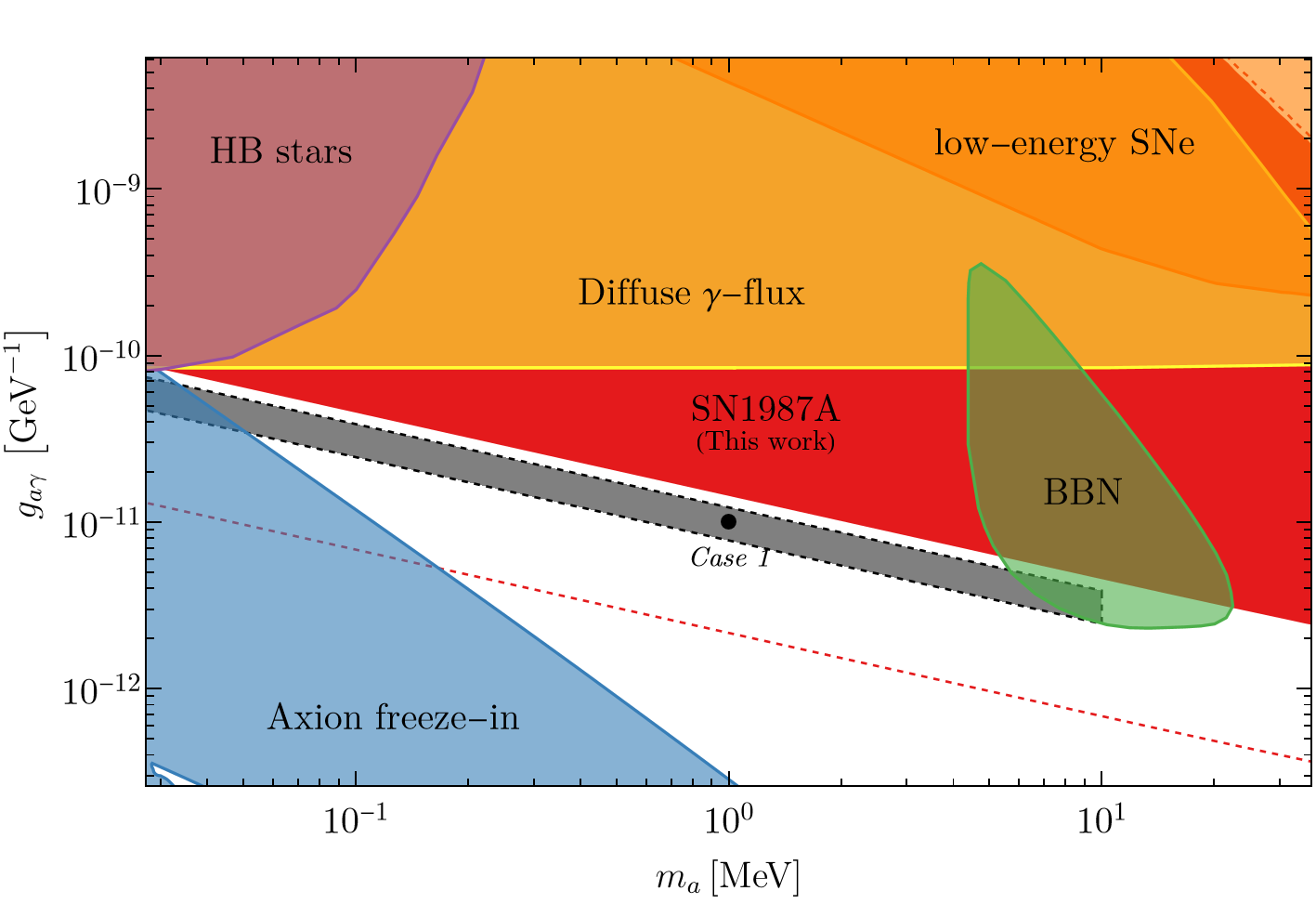}
    \caption{Best-fit values for the ALP-parameters for \textbf{case 1}. The gray region marks the area between the 16\% and the 84\% quantile of the posterior of $ \gag^2 m_a $, i.e.~between the dashed lines in the left panel of \cref{fig:ALP_posteriors_ALPparams1}. Also shown are the relevant bounds as in \cref{fig:updatedJMRbound}, and our estimate of the sensitivity of \lat~as dashed red line.}
    \label{fig:reconstructedALPparameterSpace}
\end{figure}

Fitting with our simple model, we can merely derive the posterior distribution of the product $g_{a\gamma}^2m_a$ by sampling from the joint posterior distribution $\bm{\pi}(\left.A, \omega_a^0, \alpha\right|\bm{n})$ and \cref{eq:FitParam}, and sampling from the (statistically independent) distribution of $C_0$ in \cref{eq:SN_ALP_params} for the assumed progenitor mass of $18\;M_{\odot}$. The results are provided in \cref{fig:ALP_posteriors_ALPparams1}. As a consistency check, we compare in the same figure the one-dimensional marginal posterior distributions $\bm{\pi}(\left.\omega_a^0\right|\bm{n}, A, \alpha)$ and $\bm{\pi}(\left.\alpha\right|\bm{n}, A, \omega_a^0)$ with the corresponding scatter of values found from numerical SN simulations in Ref.~\cite{Calore:2021hhn}, see \cref{eq:FitParam}. The latter comparison reveals largely consistent values.
We also show the region of best-fit values, between the 16\% and 84\% quantiles, for $ \gag $ and $ m_a $ in \cref{fig:reconstructedALPparameterSpace} along with the relevant bounds in this part of the parameter space as in \cref{fig:updatedJMRbound}. As expected, the ``correct'' value of ALP coupling and mass, i.e.~the values used to simulate the mock-observation, marked with a black dot, lies within the region. Furthermore, the entire region is contained in our \lat~sensitivity estimate, as one would expect. We cut the region off at $ m_a = 10 $~MeV since we can infer from the quality of the fit and the values of the parameters $ \alpha $ and $ \omega_a^0 $ that the ALP-mass is below 10~MeV, as we will discuss for \textbf{case 2}.
Thus, while it would not be possible to infer $ \gag $ and $ m_a $ directly from the hypothetical gamma-ray flux observation considered here, one could constrain the mass to roughly two orders of magnitude and the photon coupling to within one order of magnitude.

\begin{figure}
    \centering
    \includegraphics[width=.9\textwidth]{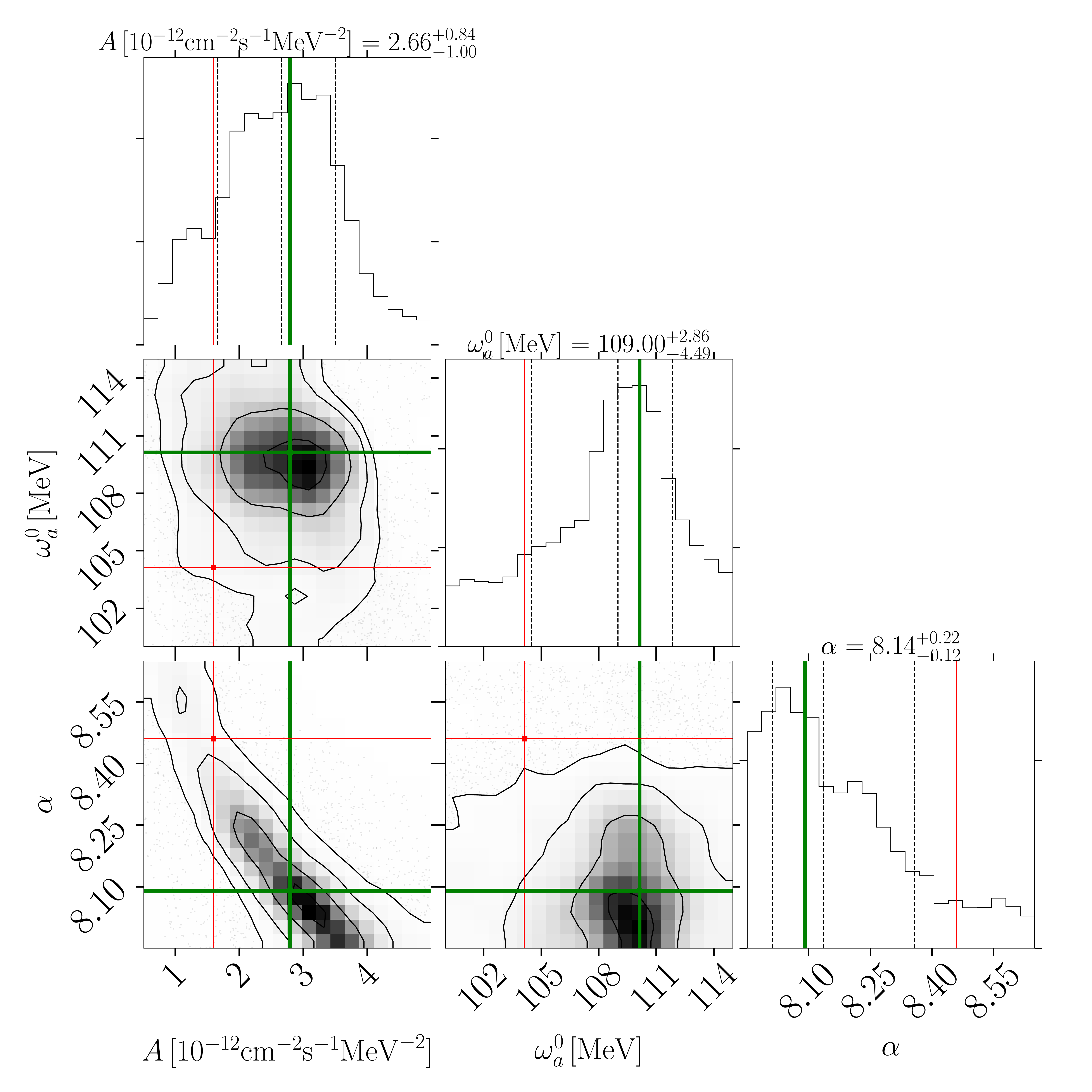}
    \caption{Same as \cref{fig:fit_alpParams1} displaying the inferred parameters for an ALP-induced gamma-ray signal from a future SN characterized by $m_a = 100$ MeV, $g_{a\gamma} = 4\times10^{-13}$ GeV$^{-1}$ (\textbf{case 2}). We employed the model in \cref{eq:alphaFitURFlux}.} 
    \label{fig:fit_alpParams2_3params}
\end{figure}

\paragraph{Case 2.}
When confronted with a signal from a SN explosion, the mass of the ALP potentially causing this signal is, of course, unknown. Consequently, it is not clear \textit{a priori} if the signal model in \cref{eq:alphaFitURFlux} is a good description of the flux.
As discussed above, the light curves of the signals (right panel of \cref{fig:ALP_spectra}) do not discriminate between the two cases (light or heavy ALPs) since no time dependence is observed.
The time-integrated spectra (left panel of \cref{fig:ALP_spectra}) also show a very similar energy dependence in the two cases.
However, the spectra still contain information about the ALP mass when accurately studied. Indeed, we show that fitting the signal with our simple, time-independent model in \cref{eq:alphaFitURFlux} yields very different results than in \textbf{case 1}.

The obtained marginal posterior distributions are shown in \cref{fig:fit_alpParams2_3params}. These results indicate that it is possible to distinguish between heavy and light ALPs. In particular, the spectral parameter $\alpha$ attains extraordinarily large values, while $A$ is very small, which we did not find when fitting the numerically determined spectrum of ALPs produced in the SN with the model in \cref{eq:alpSpectrumFit} for any mass.
Therefore these best-fit values are not compatible with the hypothesis of a light ALP ($m_a \leq 10$ MeV).
Note furthermore the large relative uncertainty on the value of $ A $ of almost $ 30\% $, reflecting the fact that the signal in \textbf{case 2} is not well described by the model.

In conclusion, in the event of a nearby SN and the observation of an associated gamma-ray signal, it is always possible to infer some information about the mass of the ALP. The spectral parameter is a good indicator to, at the very least, differentiate between different ALP mass ranges, namely those that allow for the use of \cref{eq:alphaFitURFlux} and those which do not, with the boundary between these two at around $ m_a \sim 10 \MeV $. Note that for computational purposes the prior ranges for the simulations have been chosen not wide enough to exhaustively cover and reflect the width of the posterior distributions. Because of this fact and the poor fit of the model in general, we refrain from generating a marginal posterior distribution for $ \gag $ in this case. 

Finally, in case of the detection of a signal not compatible with the light ALP hypothesis, it could still be possible to reconstruct the ALP properties by fitting the data with the complete description of this flux given by the numerical integration over \cref{eq:diffFluenceObsVars} (or the small-angle approximation in \cref{eq:diffFluenceSmallAngle} whenever applicable), and by using the numerically determined ALP spectrum.
This approach is however technically challenging since the simulation of each model photon count takes much longer and large tables have to be used for the numerically determined ALP spectra. We leave this analysis for a future investigation, as it would be warranted if indeed a gamma-ray signal following a nearby SN would be observed.

\section{Conclusions}
\label{sec:conclusions}

Heavy ALPs in the keV-MeV mass range can be very efficiently produced in core-collapse SN explosions. If they escape the progenitor's photosphere, they then decay into a burst of gamma rays that could be observed by satellite-borne detectors.
In this paper, we have conducted a comprehensive revision of this type of phenomenology, incorporating the appropriate spectrum for massive ALPs produced in a SN and developing a new, efficient method to calculate the expected gamma-ray signal. 

We extended the constraint from a gamma-ray burst caused by the decay of ALPs produced in SN 1987A to $ m_a \sim 280 \MeV $ by including the previously neglected photon coalescence process, which dominates ALP production for $ m_a \gtrsim 70 \MeV $.
Furthermore, we introduced a new form of the differential gamma-ray fluence with \cref{eq:diffFluenceObsVars}. With this \textit{observer variable} approach we proved the widely used small-angle approximation, that makes the expression for the fluence significantly simpler at early observation times --- and not only for short ALP decay lengths as previously assumed. We derived a rigorous limit on the observation time after which the small-angle approximation is not valid anymore. In such a situation, our work provides a reliable way to calculate not only the total fluence of gamma rays but also its time-dependent spectral properties.
We apply this latter method to simulate the signal that would be observed by \lat~from a future nearby SN explosion. The sensitivity, energy range and sky coverage of this detector would allow to probe ALP-photon couplings down to $ \gag \sim 10^{-13} \GeV^{-1} $ for a SN at $ \dSN = 51.4 $~kpc. If no gamma rays coincident with the SN are observed, this would yield bounds about an order of magnitude stronger in the coupling and for a factor of 2 larger ALP masses than those set by SN 1987A.

Furthermore, we have explored the possibility to reconstruct ALP properties from \lat~data in case a gamma-ray signal is indeed observed after a future SN. We showed that it is possible to characterize and give a simple prescription for the gamma-ray flux induced by the decay of light ALPs with $ m_a \lesssim 10 \MeV $. By fitting the time-independent part of the flux to the analytical approximate form predicted by our analysis, we show that the product $ \gag^2 m_a $ can be reconstructed with a small uncertainty, $\mathcal{O}(10\%)$ in the considered case. Additionally, we showed that the signal induced by ALPs heavier than $ 10 \MeV $ can be clearly recognized because the fit does not yield results compatible with the well-known ALP production mechanisms in a SN.
This conclusion was expected since the proposed analytical form for the gamma-ray signal is not valid for heavy ALPs. Thus, while we cannot reconstruct precise values for the parameters of heavy ALPs, this qualitative difference in the fitting results makes it at least possible to decide if the ALP mass is above or below $ 10 \MeV $.
A more refined study to reconstruct the heavy ALP properties is, in principle, possible thanks to the formalism we developed. However, the fitting function that we proposed is highly non-trivial, making this study technically challenging. We postpone this analysis to a future work, which is expected to give a lot of information on the ALP properties, allowing to disentangle the degeneracy between $ \gag $ and $m_{a}$.
Note also that a number of future telescopes plan to measure gamma rays in the few-MeV range, see e.g.~the list in Ref.~\cite{Carenza:2022som}, and it would be an interesting extension of this analysis to study the potential of those instruments to measure the signal of decaying ALPs from a nearby SN.
In conclusion, this work highlights that SNe are powerful ALP sources and any future nearby SN will be a unique probe of fundamental physics.

\section*{Acknowledgements}
We thank Sebastian Hoof for very useful discussions. For this work, we have made use of the \textit{Axion-Limits} repository \cite{AxionLimits}, especially for some of the bounds in \cref{fig:updatedJMRbound}.
The work of E.~M., P.~C., and D.~M.~is supported by the European Research Council under Grant No.~742104 and by the Swedish Research Council (VR) under grants  2018-03641 and 2019-02337. C.~E.~acknowledges support by the ``Agence Nationale de la Recherche'', grant n. ANR-19-CE31-0005-01 (PI: F.~Calore). The work of C.~E. has further been supported by the European Open Science Cloud (EOSC) Future project which is co-funded by the European Union Horizon Programme call INFRAEOSC-03-2020, Grant Agreement No. 101017536.
This article/publication is based upon work from COST Action COSMIC WISPers CA21106, supported by COST (European Cooperation in Science and Technology).

\bibliographystyle{bibi}
\bibliography{biblio.bib}

\providecommand{\href}[2]{#2}\begingroup\raggedright\begin{thebibliography}{10}

\bibitem{Cadamuro:2010cz}
D.~Cadamuro, S.~Hannestad, G.~Raffelt and J.~Redondo, \emph{{Cosmological
  bounds on sub-MeV mass axions}},
  \href{https://doi.org/10.1088/1475-7516/2011/02/003}{\emph{JCAP} {\bfseries
  02} (2011) 003} [\href{https://arxiv.org/abs/1011.3694}{{\ttfamily
  1011.3694}}].

\bibitem{Cadamuro:2011fd}
D.~Cadamuro and J.~Redondo, \emph{{Cosmological bounds on pseudo
  Nambu-Goldstone bosons}},
  \href{https://doi.org/10.1088/1475-7516/2012/02/032}{\emph{JCAP} {\bfseries
  02} (2012) 032} [\href{https://arxiv.org/abs/1110.2895}{{\ttfamily
  1110.2895}}].

\bibitem{Depta:2020wmr}
P.~F. Depta, M.~Hufnagel and K.~Schmidt-Hoberg, \emph{{Robust cosmological
  constraints on axion-like particles}},
  \href{https://doi.org/10.1088/1475-7516/2020/05/009}{\emph{JCAP} {\bfseries
  05} (2020) 009} [\href{https://arxiv.org/abs/2002.08370}{{\ttfamily
  2002.08370}}].

\bibitem{Balazs:2022tjl}
C.~Bal\'azs et~al., \emph{{Cosmological constraints on decaying axion-like
  particles: a global analysis}},
  \href{https://doi.org/10.1088/1475-7516/2022/12/027}{\emph{JCAP} {\bfseries
  12} (2022) 027} [\href{https://arxiv.org/abs/2205.13549}{{\ttfamily
  2205.13549}}].

\bibitem{Raffelt:1987yu}
G.~G. Raffelt and D.~S.~P. Dearborn, \emph{{Bounds on Hadronic Axions From
  Stellar Evolution}},
  \href{https://doi.org/10.1103/PhysRevD.36.2211}{\emph{Phys. Rev. D}
  {\bfseries 36} (1987) 2211}.

\bibitem{Carenza:2020zil}
P.~Carenza, O.~Straniero, B.~D\"obrich, M.~Giannotti, G.~Lucente and
  A.~Mirizzi, \emph{{Constraints on the coupling with photons of heavy
  axion-like-particles from Globular Clusters}},
  \href{https://doi.org/10.1016/j.physletb.2020.135709}{\emph{Phys. Lett. B}
  {\bfseries 809} (2020) 135709}
  [\href{https://arxiv.org/abs/2004.08399}{{\ttfamily 2004.08399}}].

\bibitem{Dolan:2021rya}
M.~J. Dolan, F.~J. Hiskens and R.~R. Volkas, \emph{{Constraining axion-like
  particles using the white dwarf initial-final mass relation}},
  \href{https://doi.org/10.1088/1475-7516/2021/09/010}{\emph{JCAP} {\bfseries
  09} (2021) 010} [\href{https://arxiv.org/abs/2102.00379}{{\ttfamily
  2102.00379}}].

\bibitem{Lucente:2022wai}
G.~Lucente, O.~Straniero, P.~Carenza, M.~Giannotti and A.~Mirizzi,
  \emph{{Constraining Heavy Axionlike Particles by Energy Deposition in
  Globular Cluster Stars}},
  \href{https://doi.org/10.1103/PhysRevLett.129.011101}{\emph{Phys. Rev. Lett.}
  {\bfseries 129} (2022) 011101}
  [\href{https://arxiv.org/abs/2203.01336}{{\ttfamily 2203.01336}}].

\bibitem{Sung:2019xie}
A.~Sung, H.~Tu and M.-R. Wu, \emph{{New constraint from supernova explosions on
  light particles beyond the Standard Model}},
  \href{https://doi.org/10.1103/PhysRevD.99.121305}{\emph{Phys. Rev. D}
  {\bfseries 99} (2019) 121305}
  [\href{https://arxiv.org/abs/1903.07923}{{\ttfamily 1903.07923}}].

\bibitem{Caputo:2022mah}
A.~Caputo, H.-T. Janka, G.~Raffelt and E.~Vitagliano, \emph{{Low-Energy
  Supernovae Severely Constrain Radiative Particle Decays}},
  \href{https://doi.org/10.1103/PhysRevLett.128.221103}{\emph{Phys. Rev. Lett.}
  {\bfseries 128} (2022) 221103}
  [\href{https://arxiv.org/abs/2201.09890}{{\ttfamily 2201.09890}}].

\bibitem{Jaeckel:2015jla}
J.~Jaeckel and M.~Spannowsky, \emph{{Probing MeV to 90 GeV axion-like particles
  with LEP and LHC}},
  \href{https://doi.org/10.1016/j.physletb.2015.12.037}{\emph{Phys. Lett. B}
  {\bfseries 753} (2016) 482}
  [\href{https://arxiv.org/abs/1509.00476}{{\ttfamily 1509.00476}}].

\bibitem{Dolan:2017osp}
M.~J. Dolan, T.~Ferber, C.~Hearty, F.~Kahlhoefer and K.~Schmidt-Hoberg,
  \emph{{Revised constraints and Belle II sensitivity for visible and invisible
  axion-like particles}},
  \href{https://doi.org/10.1007/JHEP12(2017)094}{\emph{JHEP} {\bfseries 12}
  (2017) 094} [\href{https://arxiv.org/abs/1709.00009}{{\ttfamily
  1709.00009}}]. [Erratum: JHEP 03, 190 (2021)].

\bibitem{Dobrich:2019dxc}
B.~D\"obrich, J.~Jaeckel and T.~Spadaro, \emph{{Light in the beam dump - ALP
  production from decay photons in proton beam-dumps}},
  \href{https://doi.org/10.1007/JHEP05(2019)213}{\emph{JHEP} {\bfseries 05}
  (2019) 213} [\href{https://arxiv.org/abs/1904.02091}{{\ttfamily
  1904.02091}}]. [Erratum: JHEP 10, 046 (2020)].

\bibitem{Banerjee:2020fue}
{\scshape NA64} Collaboration, D.~Banerjee et~al., \emph{{Search for Axionlike
  and Scalar Particles with the NA64 Experiment}},
  \href{https://doi.org/10.1103/PhysRevLett.125.081801}{\emph{Phys. Rev. Lett.}
  {\bfseries 125} (2020) 081801}
  [\href{https://arxiv.org/abs/2005.02710}{{\ttfamily 2005.02710}}].

\bibitem{2021ExA....51.1225D}
A.~{De Angelis}, V.~{Tatischeff}, A.~{Argan}, S.~{Brandt}, A.~{Bulgarelli},
  A.~{Bykov}, E.~{Costantini}, R.~{Curado da Silva}, I.~A. {Grenier},
  L.~{Hanlon}, D.~{Hartmann}, M.~{Hernanz}, G.~{Kanbach}, I.~{Kuvvetli},
  P.~{Laurent}, M.~N. {Mazziotta}, J.~{McEnery}, A.~{Morselli}, K.~{Nakazawa},
  U.~{Oberlack}, M.~{Pearce}, J.~{Rico}, M.~{Tavani}, P.~v. {Ballmoos},
  R.~{Walter}, X.~{Wu}, S.~{Zane}, A.~{Zdziarski} and A.~{Zoglauer},
  \emph{{Gamma-ray astrophysics in the MeV range}},
  \href{https://doi.org/10.1007/s10686-021-09706-y}{\emph{Experimental
  Astronomy} {\bfseries 51} (2021) 1225}
  [\href{https://arxiv.org/abs/2102.02460}{{\ttfamily 2102.02460}}].

\bibitem{Lucente:2020whw}
G.~Lucente, P.~Carenza, T.~Fischer, M.~Giannotti and A.~Mirizzi, \emph{{Heavy
  axion-like particles and core-collapse supernovae: constraints and impact on
  the explosion mechanism}},
  \href{https://doi.org/10.1088/1475-7516/2020/12/008}{\emph{JCAP} {\bfseries
  12} (2020) 008} [\href{https://arxiv.org/abs/2008.04918}{{\ttfamily
  2008.04918}}].

\bibitem{Giannotti:2010ty}
M.~Giannotti, L.~D. Duffy and R.~Nita, \emph{{New constraints for heavy
  axion-like particles from supernovae}},
  \href{https://doi.org/10.1088/1475-7516/2011/01/015}{\emph{JCAP} {\bfseries
  01} (2011) 015} [\href{https://arxiv.org/abs/1009.5714}{{\ttfamily
  1009.5714}}].

\bibitem{Jaeckel:2017tud}
J.~Jaeckel, P.~C. Malta and J.~Redondo, \emph{{Decay photons from the axionlike
  particles burst of type II supernovae}},
  \href{https://doi.org/10.1103/PhysRevD.98.055032}{\emph{Phys. Rev. D}
  {\bfseries 98} (2018) 055032}
  [\href{https://arxiv.org/abs/1702.02964}{{\ttfamily 1702.02964}}].

\bibitem{Hoof:2022xbe}
S.~Hoof and L.~Schulz, \emph{{Updated constraints on axion-like particles from
  temporal information in supernova SN1987A gamma-ray data}},
  \href{https://arxiv.org/abs/2212.09764}{{\ttfamily 2212.09764}}.

\bibitem{Oberauer:1993yr}
L.~Oberauer, C.~Hagner, G.~Raffelt and E.~Rieger, \emph{{Supernova bounds on
  neutrino radiative decays}},
  \href{https://doi.org/10.1016/0927-6505(93)90004-W}{\emph{Astropart. Phys.}
  {\bfseries 1} (1993) 377}.

\bibitem{Jaffe:1995sw}
A.~H. Jaffe and M.~S. Turner, \emph{{Gamma-rays and the decay of neutrinos from
  SN1987A}}, \href{https://doi.org/10.1103/PhysRevD.55.7951}{\emph{Phys. Rev.
  D} {\bfseries 55} (1997) 7951}
  [\href{https://arxiv.org/abs/astro-ph/9601104}{{\ttfamily
  astro-ph/9601104}}].

\bibitem{Brockway:1996yr}
J.~W. Brockway, E.~D. Carlson and G.~G. Raffelt, \emph{{SN1987A gamma-ray
  limits on the conversion of pseudoscalars}},
  \href{https://doi.org/10.1016/0370-2693(96)00778-2}{\emph{Phys. Lett. B}
  {\bfseries 383} (1996) 439}
  [\href{https://arxiv.org/abs/astro-ph/9605197}{{\ttfamily
  astro-ph/9605197}}].

\bibitem{Grifols:1996id}
J.~A. Grifols, E.~Masso and R.~Toldra, \emph{{Gamma-rays from SN1987A due to
  pseudoscalar conversion}},
  \href{https://doi.org/10.1103/PhysRevLett.77.2372}{\emph{Phys. Rev. Lett.}
  {\bfseries 77} (1996) 2372}
  [\href{https://arxiv.org/abs/astro-ph/9606028}{{\ttfamily
  astro-ph/9606028}}].

\bibitem{Payez:2014xsa}
A.~Payez, C.~Evoli, T.~Fischer, M.~Giannotti, A.~Mirizzi and A.~Ringwald,
  \emph{{Revisiting the SN1987A gamma-ray limit on ultralight axion-like
  particles}}, \href{https://doi.org/10.1088/1475-7516/2015/02/006}{\emph{JCAP}
  {\bfseries 02} (2015) 006} [\href{https://arxiv.org/abs/1410.3747}{{\ttfamily
  1410.3747}}].

\bibitem{Crnogorcevic:2021wyj}
M.~Crnogor\v{c}evi\'c, R.~Caputo, M.~Meyer, N.~Omodei and M.~Gustafsson,
  \emph{{Searching for axionlike particles from core-collapse supernovae with
  Fermi LAT\textquoteright{}s low-energy technique}},
  \href{https://doi.org/10.1103/PhysRevD.104.103001}{\emph{Phys. Rev. D}
  {\bfseries 104} (2021) 103001}
  [\href{https://arxiv.org/abs/2109.05790}{{\ttfamily 2109.05790}}].

\bibitem{Hoof:2021mld}
S.~Hoof, J.~Jaeckel and L.~J. Thormaehlen, \emph{{Quantifying uncertainties in
  the solar axion flux and their impact on determining axion model
  parameters}},
  \href{https://doi.org/10.1088/1475-7516/2021/09/006}{\emph{JCAP} {\bfseries
  09} (2021) 006} [\href{https://arxiv.org/abs/2101.08789}{{\ttfamily
  2101.08789}}].

\bibitem{Raffelt:1985nk}
G.~G. Raffelt, \emph{{ASTROPHYSICAL AXION BOUNDS DIMINISHED BY SCREENING
  EFFECTS}}, \href{https://doi.org/10.1103/PhysRevD.33.897}{\emph{Phys. Rev. D}
  {\bfseries 33} (1986) 897}.

\bibitem{DiLella:2000dn}
L.~Di~Lella, A.~Pilaftsis, G.~Raffelt and K.~Zioutas, \emph{{Search for solar
  Kaluza-Klein axions in theories of low scale quantum gravity}},
  \href{https://doi.org/10.1103/PhysRevD.62.125011}{\emph{Phys. Rev. D}
  {\bfseries 62} (2000) 125011}
  [\href{https://arxiv.org/abs/hep-ph/0006327}{{\ttfamily hep-ph/0006327}}].

\bibitem{Caputo:2021rux}
A.~Caputo, G.~Raffelt and E.~Vitagliano, \emph{{Muonic boson limits: Supernova
  redux}}, \href{https://doi.org/10.1103/PhysRevD.105.035022}{\emph{Phys. Rev.
  D} {\bfseries 105} (2022) 035022}
  [\href{https://arxiv.org/abs/2109.03244}{{\ttfamily 2109.03244}}].

\bibitem{Ferreira:2022xlw}
R.~Z. Ferreira, M.~C.~D. Marsh and E.~M\"uller, \emph{{Strong supernovae bounds
  on ALPs from quantum loops}},
  \href{https://doi.org/10.1088/1475-7516/2022/11/057}{\emph{JCAP} {\bfseries
  11} (2022) 057} [\href{https://arxiv.org/abs/2205.07896}{{\ttfamily
  2205.07896}}].

\bibitem{Fischer:2021jfm}
T.~Fischer, P.~Carenza, B.~Fore, M.~Giannotti, A.~Mirizzi and S.~Reddy,
  \emph{{Observable signatures of enhanced axion emission from protoneutron
  stars}}, \href{https://doi.org/10.1103/PhysRevD.104.103012}{\emph{Phys. Rev.
  D} {\bfseries 104} (2021) 103012}
  [\href{https://arxiv.org/abs/2108.13726}{{\ttfamily 2108.13726}}].

\bibitem{Landau:1975pou}
L.~D. Landau and E.~M. Lifschits, \emph{{The Classical Theory of Fields}},
  vol.~Volume 2 of \emph{Course of Theoretical Physics}. Pergamon Press,
  Oxford, 1975.

\bibitem{Kazanas:2014mca}
D.~Kazanas, R.~N. Mohapatra, S.~Nussinov, V.~L. Teplitz and Y.~Zhang,
  \emph{{Supernova Bounds on the Dark Photon Using its Electromagnetic Decay}},
  \href{https://doi.org/10.1016/j.nuclphysb.2014.11.009}{\emph{Nucl. Phys. B}
  {\bfseries 890} (2014) 17} [\href{https://arxiv.org/abs/1410.0221}{{\ttfamily
  1410.0221}}].

\bibitem{Raffelt:1996wa}
G.~Raffelt, \emph{Stars as Laboratories for Fundamental Physics: The
  Astrophysics of Neutrinos, Axions, and Other Weakly Interacting Particles},
  Theoretical Astrophysics. University of Chicago Press, 1996.

\bibitem{Ayala:2014pea}
A.~Ayala, I.~Dom\'\i{}nguez, M.~Giannotti, A.~Mirizzi and O.~Straniero,
  \emph{{Revisiting the bound on axion-photon coupling from Globular
  Clusters}}, \href{https://doi.org/10.1103/PhysRevLett.113.191302}{\emph{Phys.
  Rev. Lett.} {\bfseries 113} (2014) 191302}
  [\href{https://arxiv.org/abs/1406.6053}{{\ttfamily 1406.6053}}].

\bibitem{Diamond:2023scc}
M.~Diamond, D.~F.~G. Fiorillo, G.~Marques-Tavares and E.~Vitagliano,
  \emph{{Axion-sourced fireballs from supernovae}},
  \href{https://arxiv.org/abs/2303.11395}{{\ttfamily 2303.11395}}.

\bibitem{Langhoff:2022bij}
K.~Langhoff, N.~J. Outmezguine and N.~L. Rodd, \emph{{Irreducible Axion
  Background}},
  \href{https://doi.org/10.1103/PhysRevLett.129.241101}{\emph{Phys. Rev. Lett.}
  {\bfseries 129} (2022) 241101}
  [\href{https://arxiv.org/abs/2209.06216}{{\ttfamily 2209.06216}}].

\bibitem{Adams:2013ana}
S.~M. Adams, C.~S. Kochanek, J.~F. Beacom, M.~R. Vagins and K.~Z. Stanek,
  \emph{{Observing the Next Galactic Supernova}},
  \href{https://doi.org/10.1088/0004-637X/778/2/164}{\emph{Astrophys. J.}
  {\bfseries 778} (2013) 164}
  [\href{https://arxiv.org/abs/1306.0559}{{\ttfamily 1306.0559}}].

\bibitem{Arnett:1989tnf}
W.~D. Arnett, J.~N. Bahcall, R.~P. Kirshner and S.~E. Woosley, \emph{{SUPERNOVA
  SN1987A}},
  \href{https://doi.org/10.1146/annurev.aa.27.090189.003213}{\emph{Ann. Rev.
  Astron. Astrophys.} {\bfseries 27} (1989) 629}.

\bibitem{Mukhopadhyay:2020ubs}
M.~Mukhopadhyay, C.~Lunardini, F.~X. Timmes and K.~Zuber, \emph{{Presupernova
  neutrinos: directional sensitivity and prospects for progenitor
  identification}},
  \href{https://doi.org/10.3847/1538-4357/ab99a6}{\emph{Astrophys. J.}
  {\bfseries 899} (2020) 153}
  [\href{https://arxiv.org/abs/2004.02045}{{\ttfamily 2004.02045}}].

\bibitem{2019ascl.soft05011F}
{Fermi Science Support Development Team}, \emph{{Fermitools: Fermi Science
  Tools}},  Astrophysics Source Code Library, record ascl:1905.011, May, 2019.

\bibitem{Meyer:2016wrm}
M.~Meyer, M.~Giannotti, A.~Mirizzi, J.~Conrad and M.~A. S\'anchez-Conde,
  \emph{{Fermi Large Area Telescope as a Galactic Supernovae Axionscope}},
  \href{https://doi.org/10.1103/PhysRevLett.118.011103}{\emph{Phys. Rev. Lett.}
  {\bfseries 118} (2017) 011103}
  [\href{https://arxiv.org/abs/1609.02350}{{\ttfamily 1609.02350}}].

\bibitem{2011EPJC...71.1554C}
G.~{Cowan}, K.~{Cranmer}, E.~{Gross} and O.~{Vitells}, \emph{{Asymptotic
  formulae for likelihood-based tests of new physics}},
  \href{https://doi.org/10.1140/epjc/s10052-011-1554-0}{\emph{European Physical
  Journal C} {\bfseries 71} (2011) 1554}
  [\href{https://arxiv.org/abs/1007.1727}{{\ttfamily 1007.1727}}].

\bibitem{Calore:2021hhn}
F.~Calore, P.~Carenza, C.~Eckner, T.~Fischer, M.~Giannotti, J.~Jaeckel,
  K.~Kotake, T.~Kuroda, A.~Mirizzi and F.~Sivo, \emph{{3D template-based
  Fermi-LAT constraints on the diffuse supernova axion-like particle
  background}}, \href{https://doi.org/10.1103/PhysRevD.105.063028}{\emph{Phys.
  Rev. D} {\bfseries 105} (2022) 063028}
  [\href{https://arxiv.org/abs/2110.03679}{{\ttfamily 2110.03679}}].

\bibitem{iminuit}
H.~Dembinski, P.~Ongmongkolkul et~al., \emph{scikit-hep/iminuit},  Dec, 2020.
\newblock 10.5281/zenodo.3949207.

\bibitem{Feroz:2008xx}
F.~Feroz, M.~P. Hobson and M.~Bridges, \emph{{MultiNest: an efficient and
  robust Bayesian inference tool for cosmology and particle physics}},
  \href{https://doi.org/10.1111/j.1365-2966.2009.14548.x}{\emph{Mon. Not. Roy.
  Astron. Soc.} {\bfseries 398} (2009) 1601}
  [\href{https://arxiv.org/abs/0809.3437}{{\ttfamily 0809.3437}}].

\bibitem{Carenza:2022som}
P.~Carenza and P.~De~la Torre~Luque, \emph{{Detecting neutrino-boosted axion
  dark matter in the MeV gap}},
  \href{https://doi.org/10.1140/epjc/s10052-023-11248-w}{\emph{Eur. Phys. J. C}
  {\bfseries 83} (2023) 110}
  [\href{https://arxiv.org/abs/2210.17206}{{\ttfamily 2210.17206}}].

\bibitem{AxionLimits}
C.~O'Hare, \emph{cajohare/axionlimits: Axionlimits},
  \url{https://cajohare.github.io/AxionLimits/}, July, 2020.
\newblock 10.5281/zenodo.3932430.

\end{thebibliography}\endgroup

\end{document}